\documentclass{aa}  

\usepackage{graphicx}
\usepackage{txfonts}
\usepackage{natbib}
\bibpunct{(}{)}{;}{a}{}{,} 
\usepackage{booktabs}
\usepackage{multicol}
\usepackage{graphicx}
\usepackage{fancyhdr}
\usepackage[hidelinks,colorlinks=true,linkcolor=blue,citecolor=blue]{hyperref}
\usepackage{amsmath}	
\usepackage{amssymb}	
\usepackage[inter-unit-product=\cdot]{siunitx}
\usepackage{enumerate}
\usepackage{multirow}
\usepackage{mathtools}
\usepackage{comment}
\usepackage{longtable}
\usepackage{tabularx}
\usepackage{xcolor} 
\usepackage{ulem} 
\usepackage{soul}
\usepackage{caption}
\usepackage{cuted}

\newcommand{\Msun}{\,\mathrm{M}_\odot}

\newcommand{\MSMBH}{M_\bullet}

\newcommand{\fEdd}{\lambda_\bullet}
\newcommand{\tonde}[1]{\left(#1\right)}

\newcommand{\average}[1]{\left\langle#1\right\rangle}

\newcommand{\fastcluster}[1]{{\fontfamily{lmtt}\selectfont FastCluster}}
\newcommand{\fastclusterAGN}[1]{{\fontfamily{lmtt}\selectfont FastClusterAGN}}
\newcommand{\sevn}[1]{{\fontfamily{lmtt}\selectfont sevn}}
\newcommand{\pagn}[1]{{\fontfamily{lmtt}\selectfont pAGN}}
\newcommand{\tsunami}[1]{{\fontfamily{lmtt}\selectfont tsunami}}
\newcommand{\mcfacts}[1]{{\fontfamily{lmtt}\selectfont McFacts}}
\newcommand{\CAT}[1]{{\fontfamily{lmtt}\selectfont CAT}}
\newcommand{\arise}[1]{{\fontfamily{lmtt}\selectfont ARISE}}
\newcommand{\gwfish}[1]{{\fontfamily{lmtt}\selectfont GWFish}}

\newcommand{\orcidicon}[1]{\href{https://orcid.org/#1}{\includegraphics[width=11pt]{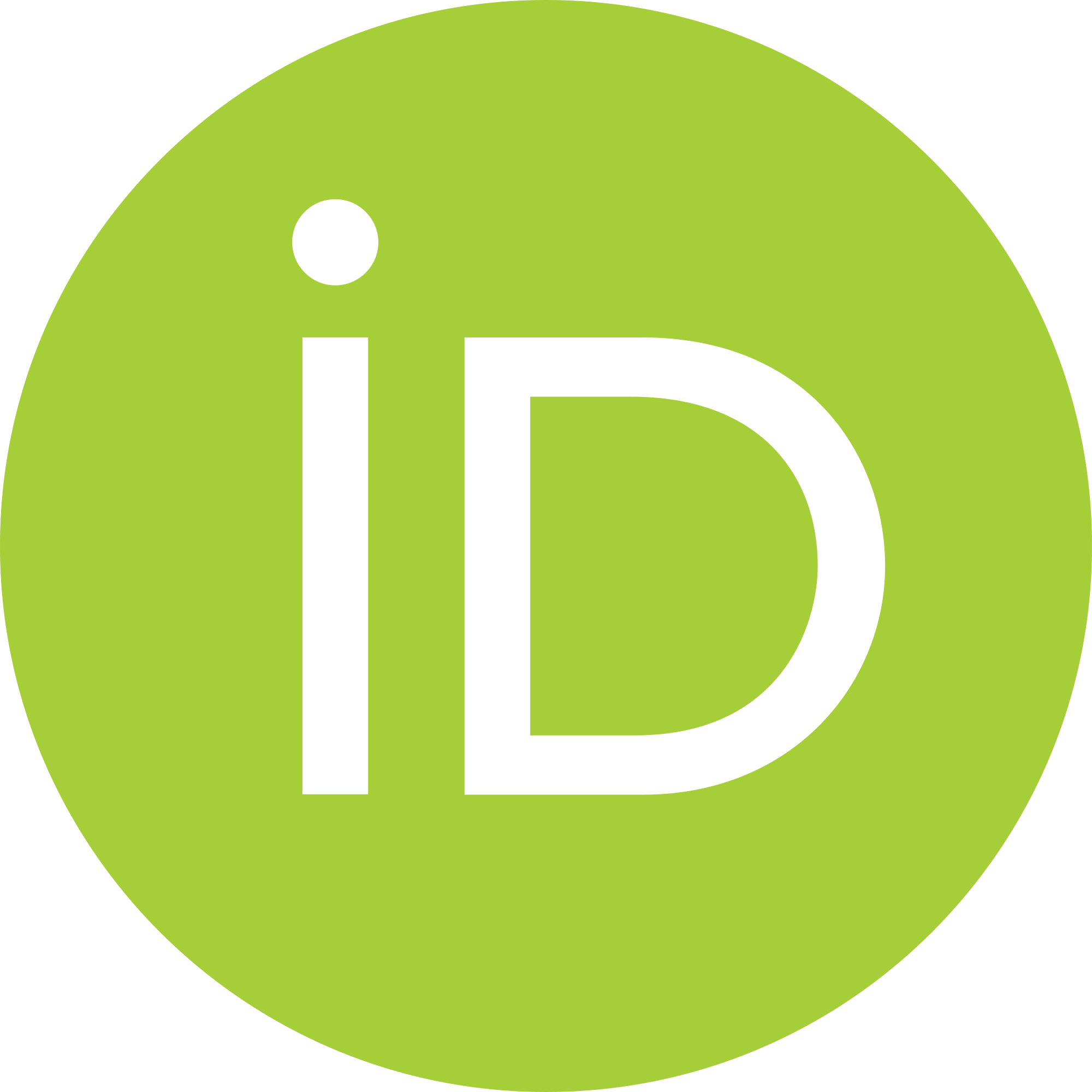}}}
\newcommand{\orcid}[1]{\href{https://orcid.org/#1}{\protect\orcidicon{#1}}}

\DeclareSIUnit\year{yr}
\DeclareSIUnit\au{AU}
\DeclareSIUnit\parsec{pc}
\DeclareSIUnit\erg{erg}

\defcitealias{SG}{SG}
\defcitealias{TQM}{TQM}
\defcitealias{Bellovary_2016}{B16}
\defcitealias{Grishin_2024}{G24}
\defcitealias{Ishibashi_2020}{IG20}
\defcitealias{Ishibashi_2024}{IG24}
\defcitealias{Calcino_2024}{C24}

\usepackage{listings}
\begin{document}

   \title{Cosmic merger rates and population evolution\\of binary black hole mergers in AGN disks}
    \titlerunning{Cosmic merger rates and population evolution of BBHs in AGN disks}

%

\author{M. Paola Vaccaro\inst{1,2}\orcid{0000-0003-3776-9246}\thanks{e-mail: \href{mailto:mariapaolavaccaro@gmail.com}{mariapaolavaccaro@gmail.com}} \and Alessandro Trinca\inst{3,4,5}\orcid{0000-0002-1899-4360} \and Michela Mapelli\inst{1,2,6,7,8}\orcid{0000-0001-8799-2548} \and  Lumen Boco\inst{1,2}\orcid{0000-0003-3127-922X} \and \\ Raffaella Schneider\inst{9,5,10,11}\orcid{0000-0001-9317-2888} \and Rosa Valiante\inst{5,10}\orcid{0000-0003-3050-1765} \and Dominika Wylezalek\inst{12}\orcid{0000-0003-2212-6045}
        }
    \authorrunning{M.~P.~Vaccaro et al.}

   \institute{Institut f{\"u}r Theoretische Astrophysik, ZAH, Universit{\"a}t Heidelberg, Albert-Ueberle-Stra{\ss}e 2, D-69120, Heidelberg, Germany
    \and
    Interdisziplin{\"a}res Zentrum f{\"u}r Wissenschaftliches Rechnen, Universit{\"a}t Heidelberg, Heidelberg, Germany
    \and
    Institute for Astronomy, University of Edinburgh, Royal Observatory, Blackford Hill, Edinburgh EH9 3HJ, UK
    \and 
    Como Lake Center for Astrophysics, DiSAT, Università degli Studi dell’Insubria, via Valleggio 11, I-22100, Como, Italy
    \and
    INAF/Osservatorio Astronomico di Roma, Via Frascati 33, I-00040 Monte Porzio Catone, Italy 
    \and
    Physics and Astronomy Department Galileo Galilei, University of Padova, Vicolo dell'Osservatorio 3, I-35122, Padova, Italy
    \and
    INAF/Osservatorio Astronomico di Padova, Vicolo dell’Osservatorio 5, I–35122, Padova, Italy
    \and 
    INFN, Sezione di Padova, Via Marzolo 8, I-35131, Padova, Italy
    \and
    Dipartimento di Fisica, “Sapienza” Università di Roma, Piazzale Aldo Moro 5, I-00185 Roma, Italy 
    \and 
    INFN, Sezione di Roma I, P.le Aldo Moro 2, I-00185 Roma, Italy 
    \and 
    Sapienza School for Advanced Studies, Viale Regina Elena 291, I-00161 Roma, Italy
    \and Astronomisches Rechen-Institut, ZAH, Universität Heidelberg, Mönchhofstr 12-14, D-69120 Heidelberg, Germany
    }

   \date{Received XXX; accepted YYY}

\abstract
{Active galactic nuclei (AGN) disks are promising sites for hierarchical binary black hole (BBH) mergers, which next-generation gravitational-wave detectors will probe out to high redshift. Predicting their contribution requires linking BBH evolution in AGN disks to the poorly constrained high-redshift evolution of active supermassive black holes (SMBHs).}
{We introduce \arise{} (AGN Rates and Intrinsic Stellar-BH Evolution), a framework that links BBH mergers in AGN disks to a cosmological description of the evolving AGN population.}
{\arise{} embeds semi-analytical models of BBH evolution in AGN disks into a cosmological description of the AGN number density, SMBH mass function, Eddington-ratio distribution, and nuclear metallicity. We consider a conservative AGN-abundance model anchored to observed luminosity functions, as well as Eddington-limited and super-Eddington scenarios based on the \CAT{} cosmological model, compatible with JWST observations of high-redshift AGNs.}
{The merger-rate density is set primarily by the disk viscosity parameter $\alpha$: at $z = 0.2$, we find a rate of $12$--$37\,\mathrm{Gpc^{-3}\,yr^{-1}}$ for $\alpha = 0.01$ and $0.5$--$6\,\mathrm{Gpc^{-3}\,yr^{-1}}$ for $\alpha = 0.1$. The redshift evolution depends strongly on how SMBHs grow. The rate density peaks at $z \simeq 1.8$, $3.3$, and $3.9$ ($1.5$, $3.6$, and $3.9$) for $\alpha=0.01$ ($0.1$) in the conservative, Eddington-limited, and super-Eddington models, respectively, and in the super-Eddington model it stays nearly flat out to $z \simeq 10$. For $\alpha = 0.1$, AGN disks account for $\sim 2$--$40\%$ of the LVK-inferred merger rate, consistent with current upper limits from GW--AGN associations. For $\alpha = 0.01$, the conservative and CAT-EL populations match or exceed it ($\sim 90$--$230\%$), while CAT-SE reaches $\sim 40$--$100\%$. The Einstein Telescope (ET) should detect $\sim 10^{5}$ ($\sim 10^{4}$) AGN-disk mergers per year for $\alpha = 0.01$ ($0.1$).}
{ET will test the AGN channel at redshifts where the AGN population models differ most strongly, providing independent constraints on the growth of early SMBHs.}


   \maketitle
   \nolinenumbers


\section{Introduction}

Supermassive black holes (SMBHs) are thought to be common in the centers of massive galaxies and to play an important role in their evolution \citep{Kormendy_Ho_2013, Heckman_2014}. During active phases of accretion, they power active galactic nuclei (AGNs), among the brightest persistent sources in the Universe. Through their intense radiation, AGNs can regulate star formation and shape the properties of their host galaxies across cosmic time \citep{Fabian_2012}.
AGN disks may also provide a favorable environment for the formation and merger of binary black holes (BBHs). Indeed, stars and stellar-origin black holes (BHs) orbiting the SMBH are subject to gas torques, which can bend their orbits, aligning them with the disk. Furthermore, BHs can migrate, pair up into binaries, and eventually merge in the dense gaseous environment surrounding the SMBH \citep{Bartos_2017, Stone_2017, Yang_2019, Secunda_2020, Tagawa_2020_b, Tagawa_2026c, Samsing_2022, Vaccaro_2023, McFacts3_2024, Dittmann_2025, Xue_2025}. 

The LIGO--Virgo--KAGRA (LVK) detector network
has now observed a large sample of BBH mergers, most recently compiled in the fifth
Gravitational-Wave Transient Catalog \citep[GWTC-5;][]{GWTC5_pop}. This sample already
enables population-level inference on the BBH merger rate and on the mass, spin, and
redshift distributions of merging BHs, which can be used to constrain
the relative contribution of different formation channels, including BBH mergers in
AGN disks. However, current detectors probe BBH mergers only out to $z\sim1.5$, leaving the high-redshift population, where AGN activity peaks
and possibly exceeds previous expectations, largely unexplored.

The next generation of ground-based gravitational-wave (GW) detectors, including the Einstein Telescope (ET) and Cosmic Explorer (CE) \citep{Punturo_2010_ET, Maggiore_2020_ET, evans2021_CE, Branchesi_2023_ET}, will dramatically extend the observable volume of the Universe, allowing BBH mergers to be detected out to redshifts of $z\sim100$ \citep{Hall_2019}. Interpreting these future observations will require models that consistently connect the BBH merger population to the cosmic evolution of AGNs and SMBHs. 

Explaining the growth and activity of SMBHs over cosmic history remains a major challenge in theoretical astrophysics \citep[e.g.,][]{Volonteri_2010}. 
This problem has become even more compelling following recent JWST observations of several populations of candidate AGNs at high redshift. These include compact red sources commonly referred to as little red dots \citep[LRDs][]{Kokorev_2023, Fan_2023, Greene_2024}, as well as AGNs selected through broad or high-ionization emission lines. Some of these candidates appear to host SMBHs with masses above $10^6 \Msun$ formed within the first billion years after the Big Bang, at redshifts as high as $z\simeq10.6$ \citep{Maiolino_2024, Bogdan_2024, Kovács_2024}. Although the physical nature of LRDs remains debated \citep{Inayoshi_2025, Matthee_2025}, number-density estimates derived from several partly independent high-redshift AGN populations suggest that active SMBHs may be substantially more abundant in the early Universe than previously assumed.
The BBH population observable by ET and CE may depend sensitively on the abundance, masses, and accretion properties of SMBHs at high redshift, potentially providing an independent probe of the growth of active SMBHs across cosmic history.

In this work, we combine a cosmological description of the AGN population with the detailed AGN-disk framework introduced by \citet{Vaccaro_2026}. There, we developed a comprehensive and up-to-date semi-analytical model for BBH formation and evolution in AGN disks, accounting for gas-driven orbital damping, migration, binary pairing, gas hardening, and three-body encounters. We embed this local model into a cosmological framework by coupling it to simulated AGN populations obtained with the semi-analytical code \CAT{} \citep[Cosmic Archaeology Tool,][]{Trinca_2022}.
Motivated by recent JWST evidence for a potentially high abundance of active SMBHs at high redshift, we explore different scenarios for the cosmological evolution of AGNs and SMBHs. Following \citet{Trinca_2022}, we consider both Eddington-limited (EL) growth  and scenarios allowing for super-Eddington (SE) accretion episodes, together with both conservative and optimistic assumptions for the abundance of high-redshift AGNs. We investigate how these different cosmological scenarios affect the BBH merger population observable by ET.

To model the joint evolution of AGNs, SMBHs, and stellar-mass BBH mergers, we introduce a dedicated cosmology wrapper, \arise{}\footnote{\arise{} is available at \href{https://gitlab.com/mariapaola.vaccaro/arise}{gitlab.com/mariapaola.vaccaro/arise}.}, which interfaces cosmological AGN populations with the AGN-disk prescriptions developed by \citet{Vaccaro_2026}. This framework allows us to consistently connect the cosmological evolution of AGNs to the properties of the BBH merger population across cosmic time.

\section{BBH mergers in AGN disks}\label{sec:fastcluster_summary}

Here, we briefly summarize our semi-analytical AGN disk population model, presented in \citet{Vaccaro_2026}. We modeled the formation and hierarchical growth of BBHs in AGN disks using the \fastcluster{} code\footnote{\fastcluster{}: \href{https://gitlab.com/micmap/fastcluster_open}{gitlab.com/micmap/fastcluster$\_$open}. \\ \fastclusterAGN{}: \href{https://gitlab.com/mariapaola.vaccaro/fastcluster_agn}{gitlab.com/mariapaola.vaccaro/fastcluster$\_$agn}.} \citep{fastcluster2021, fastcluster2022, Vaccaro_2023, Torniamenti_2024}, which follows the evolution of stellar-mass BHs embedded in AGN disks through a timescale-based approach. 

For each AGN realization, we specified the SMBH mass, $\MSMBH$, the Eddington ratio\footnote{Here, $\fEdd$ is defined as the ratio between the SMBH accretion rate, $\dot{M}_\bullet$, and the Eddington accretion rate, $\dot{M}_{\rm Edd}$.}, $\fEdd$, and the viscosity parameter of the disk, $\alpha$. We modeled the AGN disk structure using the steady-state \citet{SG} model, solved self-consistently with the \pagn{} code \citep{Gangardt_2024}. We sampled the disk lifetime from a distribution measured via the quasar proximity effect \citep{Khrykin_quasar_lifetime}.
We initialized the masses of stellar-origin BHs at a given metallicity, $Z$, using \sevn{} stellar-evolution models \citep{sevn_2023}. We modeled the stellar environment around the SMBH through an associated nuclear star cluster (NSC), whose properties scale with $\MSMBH$ \citep{Graham_Spitler, Neumayer_2020}. Their initial radial location was drawn from the steady-state profile derived by \citet{Rom_2024}, which approximates the population of BHs orbiting the SMBH before capture into the AGN disk. 

BHs initially orbiting outside the AGN disk plane are gradually captured into the disk through hydrodynamical drag. We modeled the damping of their orbital inclination using the fitting formulae of \citet{Rowan_2025}, calibrated on dedicated hydrodynamical simulations. Once embedded in the disk, BHs migrate radially under the effect of gas torques. We modeled this process using the detailed migration framework presented in \citet{Vaccaro_2025}, which follows the differential migration of BHs and determines whether two objects can converge within their mutual Hill radius, on what timescale, and at which location in the disk. In this framework, low-mass perturbers migrate according to the Type~I prescription of \citet{Grishin_2024}, while more massive BHs transition to a modified Type~II regime \citep{Kanagawa_2018}. 

Whenever two BHs come sufficiently close, we evaluate whether they can form a bound BBH, adopting the pairing criterion of \citet{Qian_2024}. The initial semimajor axis of the newly formed BBH is sampled from $p(a)\propto a^{9/2}$ between $1\,R_\odot$ and half the mutual Hill radius, while the eccentricity is drawn from a thermal distribution, $p(e)=2e$. We also allow the orbital plane of the newly-formed BBH to be either aligned or anti-aligned with the AGN disk, controlled by a free parameter, $f_\mathrm{progr}$, describing the fraction of binaries that are initially prograde. In this work, we assume $f_\mathrm{progr}=1$. The impact of different assumptions on this parameter is discussed in \citet{Vaccaro_2026}.

Once a BBH forms, we evolve it under the combined effect of gas hardening, three-body encounters, and GW emission. Gas hardening is treated following the analytical framework of \citet{Ishibashi_2024}, which is calibrated on hydrodynamical simulations of circumbinary disks. In this picture, gas torques can either shrink or widen the binary orbit depending on the local disk conditions, while also driving the evolution of the binary eccentricity. 
In addition to smooth gas-driven inspiral, BBHs may interact with other single BHs embedded in the AGN disk. We estimate the encounter timescale using the formalism of \citet{Leigh_McKernan_2018} and compare it to the inspiral timescale of the binary. If an encounter is expected before merger, we model it using a pre-computed grid of post-Newtonian three-body scattering experiments performed with the \tsunami{} code \citep{tsunami, Trani_2024}. These encounters can alter the binary semimajor axis, eccentricity, and orientation, and may lead to exchanges, ionizations, or direct collisions. At sufficiently small separations, GW emission becomes dominant, and we evolve the BBH further using the standard orbit-averaged equations of \citet{Peters_1964} until merger.
If a BBH merges, we compute the mass and spin of the remnant \citep{Jimenez, JohnsonMcDaniel_2016}, and its recoil kick velocity following \citet{maggiore_GW}. 

The remnant BH can then remain bound to the AGN disk and participate in subsequent mergers, enabling the formation of hierarchical merger chains.
We followed the evolution of each remnant using the same semi-analytical prescriptions adopted for first-generation BHs, recursively modeling its capture, migration, pairing, and inspiral. To model pairings involving higher-generation BHs, we allowed for mergers between BHs of arbitrary generations. Following \citet{Zevin-Holz}, we assumed that the probability of selecting a secondary BH of generation $M$ scales as $p(M)\propto 2^{-(M-1)}$, favoring mergers with lower-generation companions. When the companion is itself a higher-generation BH, we recursively reconstructed its mass and spin by following its previous merger history. 
Hierarchical growth continues until one of three stopping conditions is met: the cumulative merger timescale may exceed the AGN lifetime, the recoil kick imparted during a merger may eject the remnant, or the BH may simply exhaust the available supply of companions in the local AGN environment.
This treatment allowed us to self-consistently model the build-up of massive BHs through repeated mergers in AGN disks.

We sample the parameter space using a grid in $\MSMBH \in [10^5, \, 10^9]\Msun$
with a spacing of $0.1$ dex, a grid in $\fEdd\in[10^{-3}, 10]$ with a spacing
of $1$ dex, and a set of $12$ metallicity values, $Z$, in the range
$[0.01\, Z_\odot,\, Z_\odot ]$. We consider two values of the viscosity
parameter, $\alpha=0.01$ and $0.1$, and two assumptions for the dynamical
temperature of the disk, $\sigma_{\rm BH}= 0.0$ and $0.1$.\footnote{As in
\citet{Trani_2024}, we define $\sigma_{\rm BH}=\sigma_{e}=2\sigma_{i}$,
where $\sigma_{e}$ and $\sigma_{i}$ are the scale parameters of the
Rayleigh distributions of eccentricity and inclination. The case
$\sigma_{\rm BH}=0.0$ corresponds to a dynamically cold disk, whereas
$\sigma_{\rm BH}=0.1$ represents a dynamically warmer disk in which the
orbits are not fully damped.}

In \citet{Vaccaro_2026}, we carried out a systematic study of how the BBH merger efficiency, $\eta=N_{\rm merg}/N_{\rm sample}$, and the maximum remnant mass depend on $\MSMBH$, $\fEdd$, $\alpha$, and $\sigma_{\rm BH}$. We found that hierarchical growth is markedly more efficient in low-viscosity, dynamically cold disks ($\alpha = 0.01$, $\sigma_{\rm BH} = 0.0$): the higher gas surface densities of low-$\alpha$ disks shorten the capture, migration, pairing, and encounter timescales, while a fully damped BH population ($\sigma_{\rm BH} = 0.0$) further favors BBH mergers relative to the dynamically warmer, $\sigma_{\rm BH} = 0.1$ case. Within the ($\MSMBH$, $\fEdd$) plane, $\eta$ generally increases with $\fEdd$ but peaks at intermediate-to-large SMBH masses whose location shifts with $\fEdd$, tracing a band that runs from roughly $\log(\MSMBH/\Msun) \approx 6$--$7$ at $\fEdd \approx 10$ up to $\log(\MSMBH/\Msun) \approx 7$--$8$ at $\fEdd \approx 1$ \citep[see Fig.~2 in][]{Vaccaro_2026}. The most favorable configurations ($\alpha = 0.01$, $\sigma_{\rm BH} = 0.0$) within this band reach $\eta \gtrsim 70\%$ and produce remnants as massive as $\sim 2\times10^4\Msun$. Outside this region mergers are much rarer ($\eta \ll 1$) and growth saturates at far lower masses.

\section{Methods: Cosmological framework}\label{sec:methods}

\begin{figure*}
    \centering
    \includegraphics[width=\linewidth]{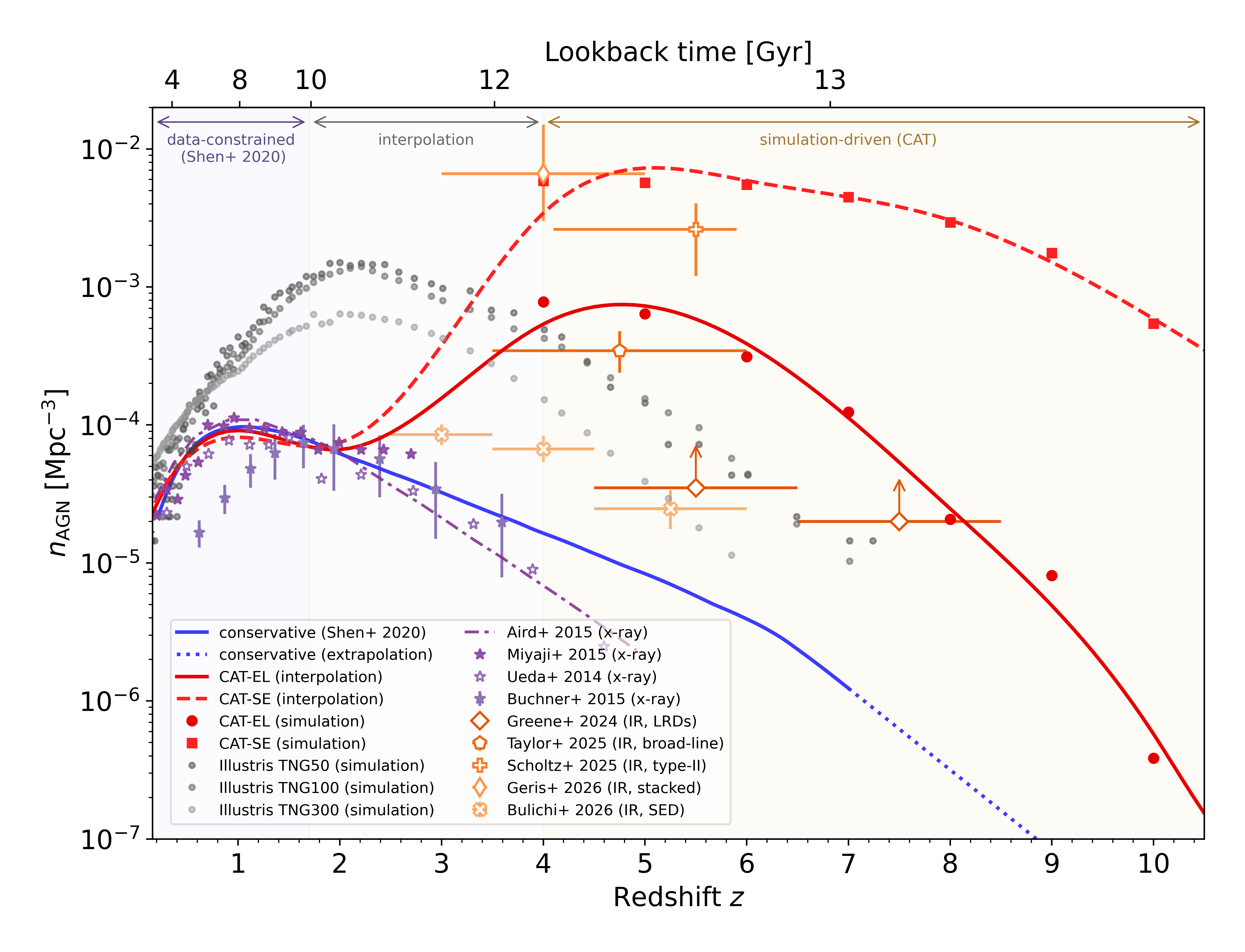}
    \caption{
Comoving AGN number density, $n_{\rm AGN}$, as a function of redshift, $z$. 
Purple symbols and curves show observational constraints from UV and X-ray AGN luminosity functions \citep{Ueda_2014, Aird_2015, Buchner_2015, Miyaji_2015, 
Shen_2020}, while orange symbols show recent high-redshift estimates based on several JWST-selected populations, including LRDs, broad-line AGNs, Type-II AGNs selected through high-ionization lines, and broad-line components inferred from stacked galaxy spectra \citep{Greene_2024, Taylor_2025, Scholtz_2025, Geris_2026, Bulichi_2026}. 
All observational estimates are selected using a bolometric luminosity threshold of $L_{\rm bol} \geq 10^{45}\, \mathrm{erg}/\mathrm{s}$.
Red symbols show the \CAT{} predictions in the Eddington-limited (circles; CAT-EL) and super-Eddington (squares; CAT-SE) scenarios \citep{Trinca_2022}. Grey points show the AGN number densities inferred from \textsc{IllustrisTNG} \citep{Illustris_TNG_1, TNG100_1, TNG100_2}.
For both simulations, AGNs are selected using a threshold of $\fEdd \geq 0.01$.  
The solid blue curve shows $n_{\rm AGN}$ for the conservative model, the dotted blue curve shows our extrapolation beyond $z=7$. The solid and dashed red curves show the corresponding interpolations for the CAT-EL and CAT-SE models, respectively.}

    \label{fig:nAGN_allz_data_comparison}
\end{figure*}

To estimate the BBH merger rate density in AGN disks across cosmic time, we develop a self-consistent treatment of the cosmological evolution of the AGN number density, SMBH mass distribution, and accretion properties as a function of redshift. 
We model the population of active SMBHs in the Universe by combining observational constraints \citep{Shen_2020} with simulated data at high redshift ($z\geq 4$) obtained with the semi-analytical code \CAT{} \citep{Trinca_2022}, as shown in \autoref{fig:nAGN_allz_data_comparison}. We use these models to reconstruct the BBH merger rate density and intrinsic BBH merger populations as a function of redshift.

\subsection{AGNs in cosmic history}

\subsubsection{Observational constraints: an evolving picture}

We discuss available observational constraints to the AGN number density in \autoref{app:review_AGN_obs}. We include results obtained from UV and X-ray surveys \citep{Ueda_2014, Aird_2015, Buchner_2015, Miyaji_2015, Shen_2020} as well as from recent JWST-based estimates \citep{Greene_2024, Taylor_2025, Scholtz_2025, Geris_2026, Bulichi_2026}. \autoref{fig:nAGN_allz_data_comparison} compares them with predictions from \CAT{} \citep{Trinca_2022} and \textsc{IllustrisTNG} \citep{Illustris_TNG_1, TNG100_1, TNG100_2} simulations. For both simulations, we compute $n_{\rm AGN}(z)$ using the activity criterion $\fEdd \geq 0.01$. 

\subsubsection{Our models: conservative, CAT-EL and CAT-SE}

We base our cosmological AGN population on the \CAT{} semi-analytic simulations \citep{Trinca_2022}, considering two variants of the \CAT{} simulation: the Eddington-limited (EL) and super-Eddington (SE) scenarios. In the EL scenario, seed SMBHs grow through spherical accretion according to the Bondi-Hoyle-Lyttleton rate, with the accretion rate capped at the Eddington limit. This imposes strong limitations on SMBH growth at high redshift. In the SE model, instead, super-Eddington accretion is allowed, typically triggered by gas-rich galaxy mergers, enabling SMBHs to grow more rapidly. We will refer to these models as ``CAT-EL'' and ``CAT-SE'', respectively. For each of these two scenarios, \CAT{} provides the SMBH mass function, $\Phi\tonde{\MSMBH, \dot{\MSMBH}, z}$, at integer
redshifts $z\geq4$ \citep{Trinca_2022}, from which we construct the  corresponding redshift-dependent SMBH mass distribution, $p(\MSMBH\mid z)$, and mass-accretion-rate distribution, $p(\fEdd \mid \MSMBH,z)$, as described in \autoref{app:M_SMBH_z} and \ref{app:fEdd_M_z}.

We estimate the fraction of active SMBHs, $f_\mathrm{active}\tonde{\MSMBH, z}$, from \CAT{} data as
\begin{equation}
f_\mathrm{active}\tonde{\MSMBH, z} =
\frac
{\int_{\dot{\MSMBH}_\mathrm{min}}^{\infty} d\dot{\MSMBH}\,
\Phi\tonde{\MSMBH, \dot{\MSMBH},z}}
{\int_{0}^{\infty} d\dot{\MSMBH}\,
\Phi\tonde{\MSMBH, \dot{\MSMBH},z}} ,
\end{equation}
assuming $\lambda_\mathrm{\bullet, min} = \dot{\MSMBH}_\mathrm{min}/\dot{\MSMBH}_\mathrm{Edd} = 0.01 $ as the threshold for a SMBH to be classified as active, as in \citet{Geris_2026}. The AGN number density can then be computed as
\begin{equation}
n_\mathrm{AGN}\tonde{z} =
\int_{10^{5} \Msun}^{10^{11} \Msun}
d\MSMBH\, \Phi\tonde{\MSMBH, z}\,
f_\mathrm{active}\tonde{\MSMBH, z}.
\end{equation}
We adopt integration limits of $10^5$ and $10^{11}\,M_\odot$, extending from the low-mass AGN population observed in dwarf galaxies \citep{Reines_2013} to the mass scale of theoretical
upper limits for black holes growing through luminous disk accretion \citep{King_2015}. In practice, \CAT{} outputs are binned in $\log (\MSMBH/\Msun)$, and the above integrals are evaluated numerically as sums over the corresponding logarithmic mass bins.

As shown in \autoref{fig:nAGN_allz_data_comparison}, the AGN number density for both the CAT-EL and CAT-SE models at $z\geq 4$ is broadly compatible with the high AGN abundances suggested by JWST-selected high-redshift sources. However, the empirical estimates\footnote{We use their global fit A, corresponding to their fiducial
redshift-evolving 
luminosity function. 
} of the AGN number density provided by \citet{Shen_2020} up to $z\simeq7$, based on a simultaneous fit to multi-wavelength quasar luminosity functions, are one to three orders of magnitude lower than the JWST estimates. Since these estimates are primarily constrained by observations of relatively luminous AGNs, generally powered by massive SMBHs, their extension toward the lower-luminosity, lower-mass population revealed by JWST is subject to considerable uncertainty, particularly at high redshift. At least part of this discrepancy may therefore reflect the different luminosity and SMBH-mass ranges probed by the two sets of observations, rather than a genuine disagreement between measurements of the same underlying population.

For this reason, we have decided to consider also a conservative model, which agrees with the results by \citet{Shen_2020} not only at low ($z<2$) but also at high redshift.  
At $z>7$, we extrapolate the \citet{Shen_2020} trend by extending the smoothed $\log n_{\rm AGN}(z)$ curve linearly in redshift, using the logarithmic slope evaluated at $z\in[6.5, 7]$. This fiducial prescription represents a conservative scenario that does not adopt the elevated high-redshift AGN abundances suggested by recent JWST observations. Rather than reconstructing an independent SMBH mass function for this model, we rescale the \citet{Shen_2020} number density onto the same $p(\MSMBH\mid z)$ and $p(\fEdd \mid \MSMBH,z)$ used for the CAT-EL model (\autoref{app:M_SMBH_z}, \autoref{app:fEdd_M_z}): the conservative and CAT-EL models therefore share the same underlying SMBH growth history, and differ only in the assumed normalization and high-redshift evolution of $n_\mathrm{AGN}(z)$. Any difference between their predictions in the following sections thus reflects differences in AGN abundance alone, not in the underlying SMBH growth model. 

For the CAT-EL and CAT-SE models, we extend $n_\mathrm{AGN}(z)$ to lower redshifts by anchoring it to the \citet{Shen_2020} estimate at $z\lesssim2$, and connect it to the corresponding \CAT{} prediction at $z\geq4$. The intermediate range $2<z<4$ is treated as a transition region, over which we perform a spline interpolation in redshift between the observational low-redshift anchor and the simulation-based high-redshift estimates. The rapid evolution of $n_{\rm AGN}$ across this interval may reflect a combination of observational and physical effects. On the observational side, the completeness of previous surveys is expected to decrease with redshift as an increasing fraction of AGNs becomes heavily obscured. On the physical side, however, the abundance of active SMBHs observed at $z\gtrsim4$ cannot be mapped directly onto the AGN population at later epochs. Whether these SMBHs continue to accrete efficiently depends on the subsequent evolution of their host galaxies, gas supply, feedback, and AGN duty cycle. Our interpolation should therefore be regarded as a phenomenological connection between the low- and high-redshift constraints, rather than as an evolutionary track for high-redshift SMBHs.
The resulting redshift evolution of $n_\mathrm{AGN}(z)$ for the conservative,
CAT-EL, and CAT-SE models is shown in \autoref{fig:nAGN_allz_data_comparison}.

\subsubsection{Metallicity evolution of galactic nuclei}
\label{sec:metallicity_models}

The metallicity evolution of nuclear stellar populations remains poorly constrained, particularly at high redshift. We therefore adopt a simple redshift-dependent prescription,
\begin{equation}
\mu_Z (z) = \left< \log\frac{Z}{Z_\odot} \right>
=
\log a_Z + z\, b_Z\,,
\end{equation}
with $Z_\odot=0.02$, and consider two limiting scenarios. For each mean metallicity, we account for an intrinsic metallicity dispersion by drawing $\log\left(Z/Z_\odot\right)$ from a normal distribution, $\mathcal{N}(\mu_Z, \sigma_Z)$, with mean $\mu_Z$ and standard deviation $\sigma_Z=0.5$ \citep{Sgalletta_2025, Boco_2026}. 

In our flat-$Z$ model, we set $a_Z=1$ and $b_Z=0$, corresponding to a mean solar metallicity at all redshifts. This choice is motivated by observations showing that the central regions of galaxies are generally more metal-rich than their outskirts \citep{Roig_2015, Ho_2015, Belfiore_2017, Neumayer_2020} and that AGN metallicities can remain high even at early cosmic times. In particular, broad-line-region studies commonly infer solar or supersolar metallicities in quasars up to $z\sim6$, with little evidence for redshift evolution \citep[e.g.][]{Xu_2018_BLR, Wang_2022_BLR}. However, these measurements probe dense gas in the vicinity of the SMBH and do not necessarily imply equally metal-rich stellar populations in the same nuclear environment, since the emitting gas may be chemically decoupled from the surrounding stars. 

We also consider an evolving-$Z$ model, with $a_Z=1.04$ and $b_Z=-0.24$, following \citet{Santoliquido_2020}. This redshift dependence is based on measurements of the metallicity evolution of damped Lyman-$\alpha$ absorbers \citep{DeCia_2018}, re-scaled to fit the average metallicity of $z\simeq 0$ galaxies from the Sloan Digital Sky Survey \citep{Gallazzi_2008}. Such measurements trace the neutral gas content of galaxies rather than their central regions, and therefore do not provide a direct estimate of the metallicity of the stellar population interacting with an AGN disk.

These two prescriptions bracket the uncertainty associated with the chemical evolution of stars in galactic nuclei. The evolving-$Z$ model assumes that nuclear metallicity follows the evolution of the average galactic gas reservoir, whereas the flat-$Z$ model assumes efficient enrichment at early cosmic times and weak subsequent evolution. Comparing the two scenarios allows us to assess how uncertainties in nuclear metallicity propagate into the mass distribution and merger properties of the BBH population.

\subsection{Source-frame merger rate density} \label{sec:R_source}

We estimate the BBH merger rate density by combining the merger populations obtained for different AGN parameters, as described in Sec.~\ref{sec:fastcluster_summary}. We first compute the merger rate per individual AGN as a function of merger delay times, and then weight these rates by the cosmological AGN population.

For each combination $(M_\bullet, \lambda_\bullet, Z)$, we sample $N_{\rm sample}=10^4$ BHs,
and we denote by $t_{\rm del}$ the delay between the onset of the AGN episode and a BBH merger.
The AGN hosts $N_{\rm BH}(M_\bullet)$ stellar-mass BHs, computed as in \autoref{app:N_BH}. 
From the simulated population, we collect the $N_{\rm merg}$ mergers, with delay times $t_{{\rm del},i}$,
$i = 1, \dots, N_{\rm merg}$. Formally, the merger rate per unit delay time of an individual AGN
is a sum of Dirac delta functions centered on the $t_{{\rm del},i}$, rescaled by
$N_{\rm BH}/N_{\rm sample}$. In practice, we evaluate it on a grid of $N_{\rm bin}$ delay-time bins, with edges $t_k$ and widths $\Delta t_k = t_{k+1} - t_k$, where $k = 1, \dots, N_{\rm bin}+1$, replacing each delta function with a top-hat function
of unit area that spans the bin containing the corresponding merger.
The merger rate produced by an individual AGN system after a delay time $t_{\rm del}$ is then
\begin{equation}
  \Gamma(t_{\rm del} \mid M_\bullet, \lambda_\bullet, Z)
  = \frac{N_{\rm BH}}{N_{\rm sample}} \sum_{k=1}^{N_{\rm bin}}
    \frac{\Pi_k(t_{\rm del})}{\Delta t_k}
    \sum_{i=1}^{N_{\rm merg}} \Pi_k(t_{{\rm del},i}) ,
  \label{eq:gamma}
\end{equation}
where $\Pi_k(x) = 1$ if $t_k \le x < t_{k+1}$ and $\Pi_k(x) = 0$ otherwise. Within each bin, $\Gamma$ is therefore constant and equal to the number
of mergers whose delay time falls in that bin, divided by $N_{\rm sample}\,\Delta t_k$ and
multiplied by $N_{\rm BH}$. Because each top-hat has unit area, the integral of $\Gamma$ over
delay times recovers the total number of mergers per sampled BH, i.e. the merger efficiency
$\eta = N_{\rm merg}/N_{\rm sample}$, multiplied by $N_{\rm BH}$:
\begin{equation}
  \int dt_{\rm del}\, \Gamma(t_{\rm del} \mid M_\bullet, \lambda_\bullet, Z)
  = \frac{N_{\rm BH}\, N_{\rm merg}}{N_{\rm sample}}
  = N_{\rm BH}\, \eta(M_\bullet, \lambda_\bullet, Z) .
  \label{eq:eta}
\end{equation}

We then average the merger rate of individual AGNs over the distributions of AGN properties at the onset redshift $z_{\rm on}$ at which the accretion episode begins:
\begin{align}
\bar{\Gamma}(t_{\rm del} \mid z_{\rm on})
=& \label{eq:gamma_avg}
\int d\log\MSMBH\, p(\log\MSMBH \mid z_{\rm on})
\\&\int d\log\fEdd\, p(\log\fEdd\mid \MSMBH, z_{\rm on})
\nonumber\\&
\int d\log Z \, p(\log Z \mid z_{\rm on})\;
\Gamma(t_{\rm del} \mid \MSMBH, \fEdd, Z).\nonumber
\end{align}
Here, $p(\log\MSMBH \mid z_{\rm on})$ is the SMBH mass distribution, obtained by renormalizing the SMBH mass function $\Phi(\MSMBH,z_{\rm on})$ over the mass range considered (\autoref{app:M_SMBH_z}), $p(\log\fEdd \mid
\MSMBH, z_{\rm on})$ is the conditional distribution of Eddington ratios at fixed SMBH mass and redshift (\autoref{app:fEdd_M_z}), and $p(\log Z \mid
z_{\rm on})$ is the nuclear metallicity distribution (Sec.~\ref{sec:metallicity_models}). The quantity $\bar{\Gamma}(t_{\rm del}\mid z_{\rm on})$ is thus the mean merger rate, per unit delay time, of a single AGN that became active at $z_{\rm on}$.

We adopt a flat $\Lambda$CDM cosmology \citep{Planck2018}, and denote by $t(z)$ the cosmic time at redshift $z$. A merger with delay time $t_{\rm del}$ in an AGN that became active at $z_{\rm on}$ occurs at the cosmic time $t_{\rm merg} = t(z_{\rm on}) + t_{\rm del}$, and therefore at the redshift $z_{\rm merg}$ defined by $t(z_{\rm merg})=t_{\rm merg}$. 

The comoving merger rate density at $z=z_{\rm merg}$ is therefore obtained by weighting $\bar{\Gamma}$ by the comoving number density of active AGNs and summing over all onset times:
\begin{equation}
\mathcal{R}_\mathrm{AGN}(z)
=
\int_{t(z_{\rm max})}^{t(z)} dt_{\rm on}\,
\frac{d n_{\rm AGN}(z_{\rm on})}{dt_{\rm on}}\,
\bar{\Gamma}\!\left(t(z)-t_{\rm on} \mid z_{\rm on}\right),
\label{eq:R_source}
\end{equation}
where we take $z_{\rm max}=15$. 
In practice, we evaluate Eq.~\ref{eq:R_source} on the
same delay-time bins $[t_k, t_{k+1})$ used to tabulate $\bar{\Gamma}$, as
\begin{equation}
\mathcal{R}_{\rm AGN}(z) \simeq \sum_{k} n_{\rm AGN}(z_{{\rm on},k})\,
\bar{\Gamma}(t_{{\rm del},k} \mid z_{{\rm on},k}) ,
\end{equation}
where $t_{{\rm del},k}$ is the midpoint of the $k$-th bin,
$z_{{\rm on},k}$ is the onset redshift satisfying
$t(z_{{\rm on},k}) = t(z) - t_{{\rm del},k}$, and only bins with
$t_{{\rm del}} < t(z) - t(z_{\rm max})$ contribute.

\begin{figure*}
    \centering
    \includegraphics[width=\linewidth]{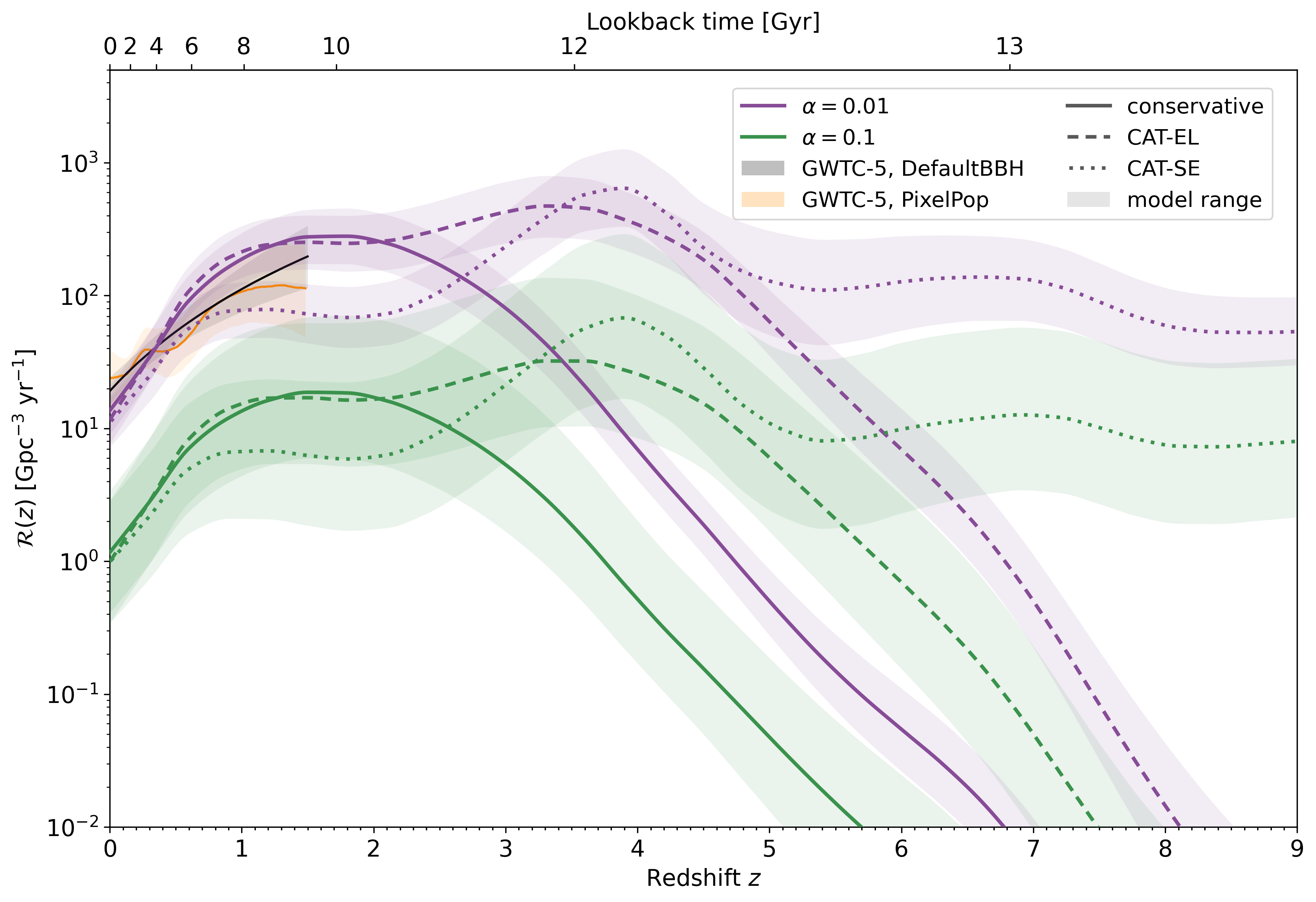}
    \caption{Source-frame BBH merger rate density, $\mathcal{R}(z)$, formed in AGN disks as a function of merger redshift. Purple and green curves correspond to disk-viscosity parameters $\alpha=0.01$ and $0.1$, respectively. Solid, dashed, and dotted lines identify the conservative, CAT-EL, and CAT-SE prescriptions for the cosmological AGN population. The central curves show the logarithmic mean over the adopted metallicity prescriptions and initial BH orbital dispersions, and the shaded bands enclose the corresponding model range. Curves are smoothed in $\log\mathcal{R}$ for display (Gaussian kernel, $\sigma_z = 0.25$–$0.45$); values quoted in the text refer to the raw curves. For comparison, we show the GWTC-5 merger rate density inferred with the DefaultBBH and PixelPop population models \citep{GWTC5_pop}. The top axis indicates the corresponding lookback time. 
}
    \label{fig:MRD(z)}
\end{figure*}

\subsection{Observed merger rate density and mock catalogs}
\label{sec:R_obs}

The procedure outlined in Sec.~\ref{sec:R_source} yields the comoving merger rate density per unit source-frame time. To convert this quantity to the observer frame, we account for cosmological time dilation. The observed merger rate per unit redshift is therefore
\begin{equation}
\frac{d\mathcal{R}_\mathrm{obs, AGN}}{dz}(z) =
\frac{\mathcal{R}_\mathrm{AGN}(z)}{1+z}\,
\frac{dV_c}{dz}(z)\,,
\label{eq:merger_rate_obv}
\end{equation}
where the factor $(1+z)^{-1}$ converts source-frame time intervals into observer-frame time intervals and $dV_c/dz$ is the comoving volume element. Hence, the expected number of intrinsic observer-frame mergers in a given redshift bin is
\begin{equation}
\left\langle N_j \right\rangle =
T_{\rm obs}
\int_{z_j-\delta z_j/2}^{z_j+\delta z_j/2}
d\mathcal{R}_\mathrm{obs, AGN}(z)
\label{eq:N_merger_obs}
\end{equation}
where $T_{\rm obs}$ is the observing time window. We denote the $j$-th redshift bin by $z_j \pm \delta z_j/2$, with $z_j$ its center and $ \delta z_j$ its width.

We construct mock observed catalogs of BBH mergers for an observing time $T_{\rm obs}=1\,{\rm yr}$. For each redshift bin, we construct the corresponding merger population, $p(\theta \mid z_j)$, from the same weighted population entering Eq.~\ref{eq:gamma_avg}
, marginalizing over formation redshift, metallicity, SMBH mass, and Eddington ratio. Here, $\theta$ denotes the intrinsic binary parameters, including the component masses and spins.
We then draw the number of intrinsic mergers in the bin, $N_j$, from a Poisson distribution with mean $\langle N_j\rangle$, and sample $N_j$ events from $p(\theta \mid z_j)$. The resulting list of events constitute the intrinsic catalog of BBH mergers during the observing time $T_{\rm obs}$, expressed in the observer frame.

We apply detector selection for the ET at the event level. 
For each merger in the intrinsic catalog, we randomly select the extrinsic parameters by sampling the binary inclination, sky position, polarization angle, and coalescence phase from isotropic distributions: $\cos\theta_{\rm JN}\sim\mathcal{U}(-1,1)$, $\alpha\sim\mathcal{U}(0,2\pi)$, $\sin\delta\sim\mathcal{U}(-1,1)$, $\psi\sim\mathcal{U}(0,\pi)$, $\phi_c\sim\mathcal{U}(0,2\pi)$, where $\theta_{\rm JN}$ is the inclination angle, $\alpha$ and $\delta$ are the right ascension and declination, $\psi$ is the polarization angle, and $\phi_c$ is the coalescence phase. We then compute the network signal-to-noise ratio (SNR), $\rho$, with \gwfish{} \citep{Dupletsa_2023}. 
We use the noise power spectral densities of the triangular (delta) configuration\footnote{We repeated the analysis assuming a configuration of two L-shaped ET detectors instead of the triangular one, and obtained similar results, both in the total number of detections and in their redshift distribution. We therefore only show the delta configuration here.} with $10\,{\rm km}$ arms for the ET and assume a detector duty cycle (i.e., the fraction of time the detector is actively taking data) of $0.85$. Waveforms are generated with the \texttt{IMRPhenomXPHM} model, which includes precession and higher-order modes \citep{Garcia_Quiros_2020, Pratten_2021}.

\section{Results}
\label{sec:results}

\subsection{Redshift evolution of the merger rate density}
\label{sec:results_MRD}

\autoref{fig:MRD(z)} shows the source-frame BBH merger rate density predicted by our AGN models as a function of merger redshift. The overall normalization depends strongly on the AGN disk viscosity. For the same cosmological prescription, models with $\alpha=0.01$ produce merger rate densities approximately one order of magnitude higher than those with $\alpha=0.1$ across most of the explored redshift range. This difference reflects the greater merger efficiency of denser, less viscous AGN disks, in which stronger gas interactions facilitate the capture and hardening of stellar-origin BHs. The disk structure also regulates BH migration and therefore the efficiency of binary pairing, both through convergence toward migration traps and through differential migration outside them \citep[see][for a detailed analysis]{Vaccaro_2026}. For each value of $\alpha$ and cosmological prescription, the central line represents the logarithmic mean of the merger rate density obtained from the four combinations of the two metallicity models described in Sec.~\ref{sec:metallicity_models} and the two initial BH orbital dispersions introduced in Sec.~\ref{sec:fastcluster_summary}. The corresponding shaded region shows the full modeling range spanned by these four predictions at each redshift.

In the conservative cosmological model, the merger rate density increases from $z\simeq0$ to a maximum at $z\simeq1.8$ ($1.5$) for $\alpha=0.01$ ($0.1$). The merger rate has a local value $\mathcal{R}(z=0.2)=24^{+13}_{-8}\,{\rm Gpc}^{-3}{\rm yr}^{-1}$ ($2.0^{+4.0}_{-1.3}\,{\rm Gpc}^{-3}{\rm yr}^{-1}$) and reaches a maximum of $280^{+170}_{-110}\,{\rm Gpc}^{-3}{\rm yr}^{-1}$ ($19^{+50}_{-13}\,{\rm Gpc}^{-3}{\rm yr}^{-1}$) for $\alpha=0.01$ ($0.1$)\footnote{Error bars indicate model ranges, not confidence intervals.}. Beyond the peak, the rate decreases rapidly: the central predictions fall below $10^{-2}\,{\rm Gpc}^{-3}{\rm yr}^{-1}$ by $z\simeq7$ ($z\simeq6$) for $\alpha=0.01$ ($0.1$). 

In the CAT-EL model, the merger rate density has similar local values of $\mathcal{R}(z=0.2)=23^{+12}_{-8}\,{\rm Gpc}^{-3}{\rm yr}^{-1}$ ($1.9^{+3.7}_{-1.3}\,{\rm Gpc}^{-3}{\rm yr}^{-1}$) for $\alpha=0.01$ ($0.1$).
However, it continues to rise until $z\simeq3.3$ ($3.6$) for $\alpha=0.01$ ($0.1$), reaching maximum values of $490^{+340}_{-210}\,{\rm Gpc}^{-3}{\rm yr}^{-1}$ for $\alpha=0.01$ and $34^{+110}_{-23}\,{\rm Gpc}^{-3}{\rm yr}^{-1}$ for $\alpha=0.1$. It then decreases steeply, similarly to the conservative model, falling below $10^{-2}\,{\rm Gpc}^{-3}{\rm yr}^{-1}$ by $z\simeq8$ ($z\simeq7.5$) for $\alpha=0.01$ ($0.1$). The conservative and CAT-EL models share the same $p(\MSMBH \mid z)$ and $p(\fEdd \mid \MSMBH, z)$, so any difference in their predicted merger rate densities reflects differences in the assumed AGN abundance, $n_{\rm AGN}(z)$.

The CAT-SE prescription predicts $\mathcal{R}(z=0.2)=18^{+10}_{-6}\,{\rm Gpc}^{-3}{\rm yr}^{-1}$ ($1.6^{+3.2}_{-1.1}\,{\rm Gpc}^{-3}{\rm yr}^{-1}$) for $\alpha=0.01$ ($0.1$), and produces a maximum in merger rate density around $z\simeq3.9$, with maximum rates of $790^{+760}_{-380}\,{\rm Gpc}^{-3}{\rm yr}^{-1}$ and $82^{+270}_{-61}\,{\rm Gpc}^{-3}{\rm yr}^{-1}$ for the low- and high-viscosity models, respectively. Unlike the CAT-EL and conservative predictions, however, the CAT-SE merger rate density does not decline at still higher redshifts. Instead, it approaches a broad plateau at $z\gtrsim6$, with $R(z)\sim82^{+77}_{-40}\,{\rm Gpc}^{-3}{\rm yr}^{-1}$ for $\alpha=0.01$ and $\sim9^{+32}_{-7}{\rm Gpc}^{-3}{\rm yr}^{-1}$ for $\alpha=0.1$, extending up to the highest redshifts considered of $z\simeq 10$.

\medskip

The merger rate density results from the convolution of the AGN abundance with the redshift-dependent SMBH-mass and Eddington-ratio distributions, hence it both traces the evolution of the AGN number density shown in \autoref{fig:nAGN_allz_data_comparison} and probes the accretion history of SMBHs across cosmic time.
As $p(\MSMBH \mid z)$ and $p(\fEdd \mid \MSMBH, z)$ evolve, they shift the
relative weights assigned to AGN disks with different properties.
As mentioned in Sec.~\ref{sec:fastcluster_summary} and shown by \citet{Vaccaro_2026}, the merger efficiency $\eta$ is maximal for intermediate-to-large SMBH masses, $6.5 \lesssim \log(\MSMBH/\Msun) \lesssim 8$, accreting at near-Eddington or super-Eddington rates, $\fEdd \gtrsim 1$.

This dependency explains the different behavior of the merger rate density predicted by the two \CAT{}-based scenarios. In the CAT-SE model, SMBHs assemble earlier: as shown in \autoref{fig:pfedd_CAT_comparison}, they undergo episodes of super-Eddington accretion predominantly at $z \gtrsim 5$, so a large fraction of the population enters the high-$\eta$ region of the parameter space at early cosmic times, which drives the plateau seen in the merger rate density at high redshift (\autoref{fig:MRD(z)}). In the CAT-EL model, instead, the accretion is strictly Eddington-limited ($\fEdd < 1$) at $z \gtrsim 4$, so SMBHs are
not allowed to grow through super-Eddington episodes this early.
In the redshift range where we connect the CAT-EL model and the \citet{Shen_2020} observational fit ($z=2-4$) we must relax the assumption of strictly Eddington-limited
accretion: to reproduce the local SMBH population, the CAT-EL model
requires a population of systems accreting at $\fEdd \gtrsim 1$ around
$z \simeq 3$ (see \autoref{app:fEdd_M_z}), rather than remaining
Eddington-limited down to low redshift. These BHs are effectively
catching up later in cosmic history than their CAT-SE counterparts: having been prevented from growing through super-Eddington accretion at $z \gtrsim 4$, the CAT-EL retains a larger population of active SMBHs with $\fEdd \gtrsim 1$ at $z \simeq 2$--$3$ than CAT-SE. Because these near- and super-Eddington systems are more efficient at producing BBH mergers, the merger rate density in the CAT-EL and
conservative models exceeds that of CAT-SE at low redshift $z\lesssim3$.

\subsection{Redshift evolution of the intrinsic BBH population}
\label{sec:results_intrinsic_population}

\begin{figure*}
    \centering
    \includegraphics[width=\linewidth]{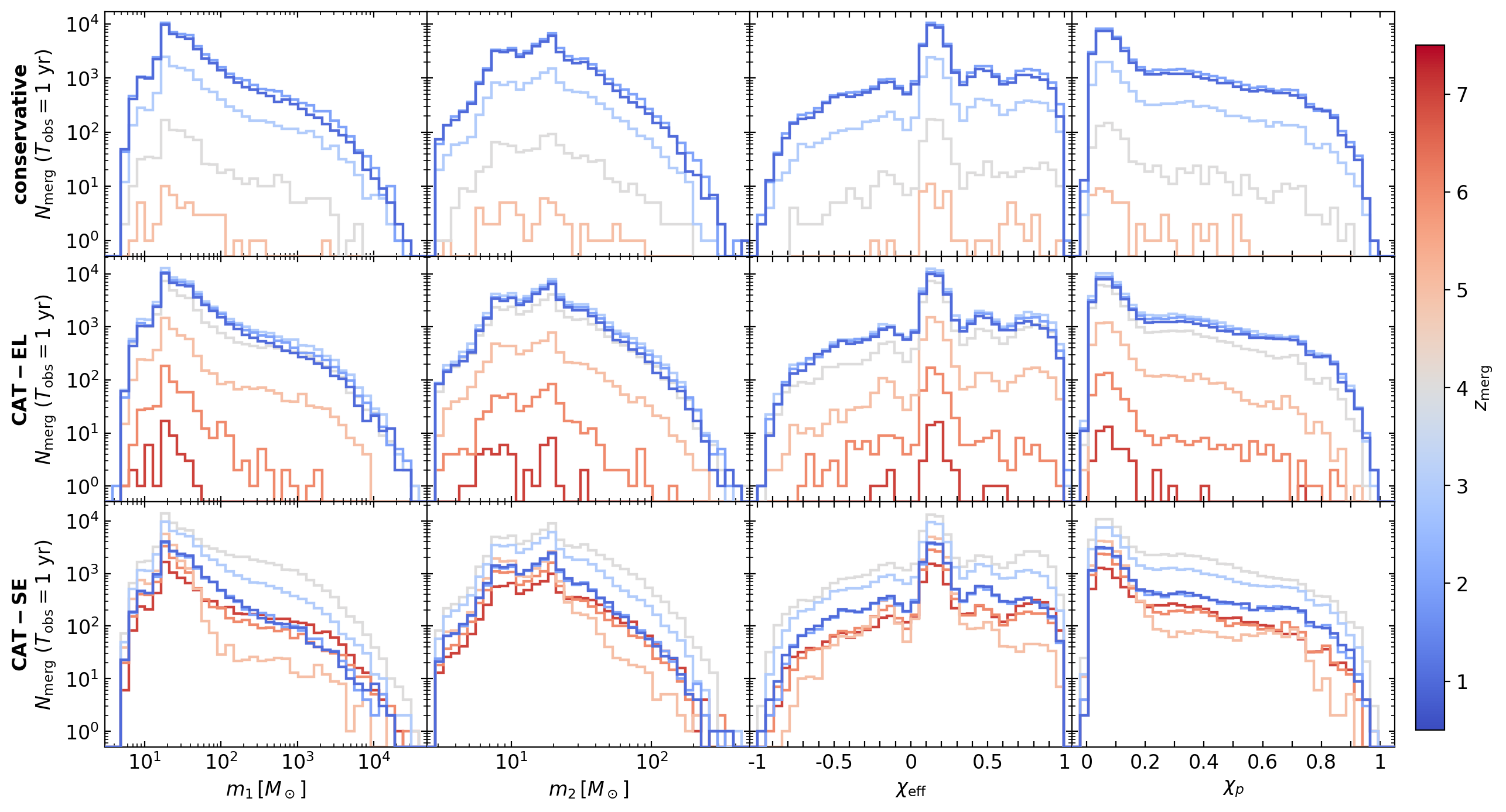}
    \caption{Merger-count distributions of BBH merger properties for three cosmological model variants: conservative (top row), CAT-EL (middle row), and CAT-SE (bottom row). We assume $\alpha=0.01$, $\sigma_{\rm BH}=0.0$, and the flat-$Z$ distribution. Columns show the primary mass, $m_1$, secondary mass, $m_2$, effective spin, $\chi_{\rm eff}$, and precessing-spin parameter, $\chi_{\rm p}$. Each curve shows the number of intrinsic mergers expected over an observing time of $T_{\rm obs}=1\,{\rm yr}$, obtained by integrating the observer-frame merger rate over redshift bins of width $\delta z = 0.5$. Colors encode the merger redshift, $z_{\rm merg}$, increasing from low redshift (blue) to high redshift (red).}
    \label{fig:redshift_colored_distributions}
\end{figure*}

We now investigate how the intrinsic properties of BBH mergers evolve across cosmic time. 
\autoref{fig:redshift_colored_distributions} shows the merger-count distributions of the primary mass, $m_1$, secondary mass, $m_2$, effective spin, $\chi_{\rm eff}$, and precessing-spin parameter, $\chi_{\rm p}$, in different merger-redshift bins. 
Here we fix the AGN-disk and stellar-population assumptions to $\alpha=0.01$, $\sigma_{\rm BH}=0.0$, and the flat-$Z$ model, and compare the conservative, CAT-EL, and CAT-SE prescriptions for the cosmological AGN population.
Each histogram represents the number of intrinsic mergers expected during an observer-frame interval of $T_{\rm obs}=1\,{\rm yr}$ in redshift bins of width $\delta z=0.5$, as in Eq.~\ref{eq:N_merger_obs}. 
Unlike the merger-rate density shown in \autoref{fig:MRD(z)}, the normalization of these distributions also includes the comoving volume accessible in each redshift shell and the cosmological time-dilation factor (see Eq.~\ref{eq:merger_rate_obv}).

The overall evolution in the number of mergers closely reflects the differences between the three cosmological prescriptions discussed in Sec.~\ref{sec:results_MRD}. 
In the conservative model, the merger population is strongly concentrated at low and intermediate redshift, $z\lesssim3$, with the number of events decreasing rapidly toward high $z$. 
The CAT-EL model retains a substantial population up to larger redshifts, $z\simeq5$, consistent with the later peak of its merger rate density. 
The contrast is strongest for CAT-SE: mergers remain numerous even in the highest-redshift bins, $z\simeq7$, shown in \autoref{fig:redshift_colored_distributions}, reflecting the sustained high-redshift activity produced by the early growth of active SMBHs in this scenario. 

Despite the large differences in normalization, the qualitative shapes of the binary-property distributions remain broadly similar across redshift.
The component-mass distributions are dominated by stellar-mass BHs at a few tens of solar masses, with an extended high-mass tail in the primary mass produced by hierarchical growth in the AGN disk. In the most populated redshift bins, the primary-mass distribution extends to $\gtrsim 10^4\,{\rm M_\odot}$. Toward higher redshift, these high-mass tails become progressively less populated in the conservative and CAT-EL cases, and eventually disappear from the histograms as the total merger count decreases.

The secondary BH mass distribution is restricted below $\sim{200}$ M$_\odot$ and declines more rapidly toward high masses. As discussed by \citet{Vaccaro_2026}, this is a characteristic consequence of the hierarchical-merger prescription: while the primary BH can grow through repeated mergers, higher-generation primaries preferentially pair with lower-generation companions, reflecting the decreasing abundance of high-generation BHs in the disk. The secondary mass therefore saturates much more rapidly, 
and hierarchical growth progressively produces more unequal-mass binaries.

The spin distributions likewise remain broad over cosmic time. The $\chi_{\rm eff}$ distribution spans both positive and negative values but shows substantial support at positive $\chi_{\rm eff}$, a feature that carries information about the geometrical orientation of BBHs at formation \citep[see][]{Vaccaro_2026}. The precessing-spin parameter $\chi_{\rm p}$ also extends over most of its allowed range. At the extremes of the spin distributions, in particular $|\chi_{\rm eff}|\gtrsim0.8$, the populations become progressively sparser and eventually disappear from the highest-redshift bins in the conservative and CAT-EL cases, reflecting the smaller number of mergers sampled at high redshift.

\autoref{fig:redshift_colored_distributions} illustrates these trends for a representative configuration with $\alpha=0.01$, $\sigma_{\rm BH}=0.0$ and the flat-$Z$ assumption, while isolating the effect of the cosmological AGN prescription. Variations of this figure for different choices of $\alpha$, $\sigma_{\rm BH}$ and metallicity evolution model are available in the supplementary material. The effects of the parameters $\alpha$ and $\sigma_{\rm BH}$ are discussed in the companion paper by \citet{Vaccaro_2026}, and the impact of the assumed metallicity evolution of galactic nuclei is shown in \autoref{app:population_variants}.

\subsection{Predictions for the Einstein Telescope (ET)}

\begin{figure*}
    \centering
    \includegraphics[width=\linewidth]{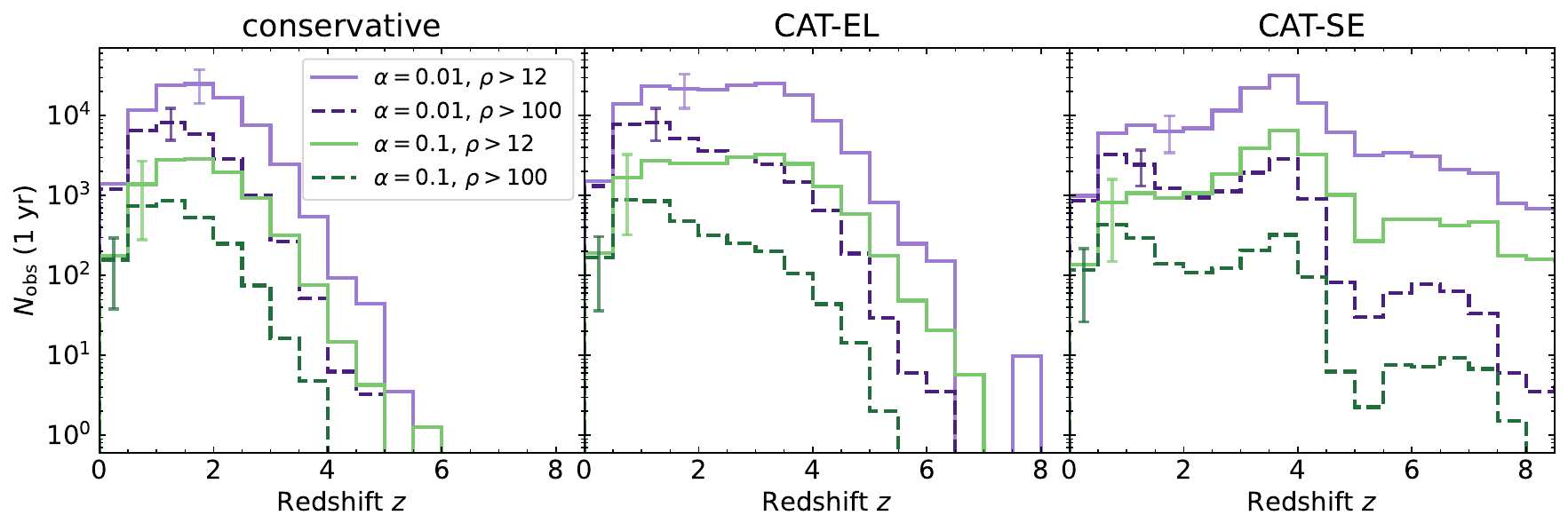}
    \caption{Expected number of BBH mergers detected by ET over an observing time of $T_{\rm obs}=1\,{\rm yr}$, as a function of merger redshift (in bins of $\delta z = 0.5$), for the conservative (left), CAT-EL (middle), and CAT-SE (right) cosmological AGN models. Purple and green lines correspond to disk viscosity parameters $\alpha=0.01$ and $\alpha=0.1$, respectively. Solid lines show the detectable population with SNR $\rho>12$, and darker dashed lines show the high-SNR subset with $\rho>100$. Lines show the logarithmic mean over the different choices of $\sigma_{\rm BH}$ and metallicity evolution model. For each curve, a representative error bar in one redshift bin shows the full range (minimum to maximum) across these realizations; the error bars are placed in different bins for readability.
}
    \label{fig:gwfish_event_counts}
\end{figure*}

\begin{table}[]
    \centering
    \begin{tabular}{llrrr}
        \hline
        \hline
        Model & $\alpha$ & $N_{\rm merg}$ & $N_{\rm obs}$ & $N_{\rm obs}$ \\
         & & & $\rho>12$ & $\rho>100$ \\
        \hline
    conservative & 0.01 & $1.7\times10^5$ & $\mathbf{8.9\times10^4}$ & $2.6\times10^4$ \\
    CAT-EL       & 0.01 & $3.4\times10^5$ & $\mathbf{1.6\times10^5}$ & $3.4\times10^4$ \\
    CAT-SE       & 0.01 & $2.9\times10^5$ & $\mathbf{1.3\times10^5}$ & $1.6\times10^4$ \\
    conservative & 0.1 & $1.5\times10^4$ & $\mathbf{1.1\times10^4}$ & $2.6\times10^3$ \\
    CAT-EL       & 0.1 & $3.5\times10^4$ & $\mathbf{2.1\times10^4}$ & $3.3\times10^3$ \\
    CAT-SE       & 0.1 & $4.2\times10^4$ & $\mathbf{2.3\times10^4}$ & $1.9\times10^3$ \\
        \hline
    \end{tabular}
    \caption{Number of BBH mergers in AGN disks expected to be detected by the ET over $T_{\rm obs}=1\,{\rm yr}$, for different cosmological AGN models and disk viscosity parameters $\alpha$. Columns report the total number of mergers, $N_{\rm merg}$, 
    the number of events with SNR $\rho>12$ (detectable population) and $\rho>100$ (high-SNR subset). These estimates are based on a single realization of the extrinsic parameters for each event, and should therefore be regarded as indicative. }
    \label{tab:gwfish_Nevents}
\end{table}

We study how the BBH populations predicted by our AGN models would be observed by next-generation ground-based GW detectors, such as the ET \citep{Branchesi_2023_ET}. We use the Fisher-matrix code \texttt{GWFish} \citep{Dupletsa_2023} to compute the network SNR, $\rho$, of each event in our mock catalogs.  
We classify an event as detectable if its network SNR exceeds a threshold $\rho > \rho_\ast$, and adopt $\rho_\ast=12$, following \citet{Muccillo_2026}. 
 
\autoref{fig:gwfish_event_counts} shows the expected number of mergers detected by ET in one year, assuming a duty cycle of 0.85, as a function of merger redshift. For each cosmological AGN model and disk viscosity parameter, we show the detectable population, with $\rho>12$, and the high-SNR subset, with $\rho>100$. The integrated number of events in these two detected populations, as well as the number of events in the intrinsic catalog, are listed in \autoref{tab:gwfish_Nevents}. 
ET is expected to detect $0.9$--$1.6\times10^5$ AGN-disk mergers with $\rho>12$ over $T_{\rm obs}=1\,{\rm yr}$ for models with $\alpha=0.01$, whereas for $\alpha=0.1$ the numbers are lower by a factor $\sim 6$--$8$, consistent with the lower merger rate densities. The cosmological AGN prescription mainly changes the redshift distribution of the detections. In the conservative model the detectable events are concentrated at $z\lesssim 3$, and the distribution drops to zero by $z\simeq 5$--$6$. The CAT-EL model extends the detections to $z\simeq 6$--$7$, whereas the CAT-SE model retains a significant number of detections up to the highest redshifts, $z\simeq 8$, even for the high-SNR subset. The high-SNR events ($\rho>100$) make up about $25-30\%$, $15-20\%$, and $10\%$ of the detectable population in the conservative, CAT-EL, and CAT-SE models, respectively, with slightly lower fraction for $\alpha=0.1$ than for $\alpha=0.01$. The smaller fraction in CAT-SE reflects the larger share of mergers at high redshift, which are more distant and therefore have lower SNR.

\section{Discussion}

\subsection{Constraints on the AGN contribution to the LVK catalog}
\label{sec:fAGN}
\begin{figure}
    \centering
    \includegraphics[width=\linewidth]{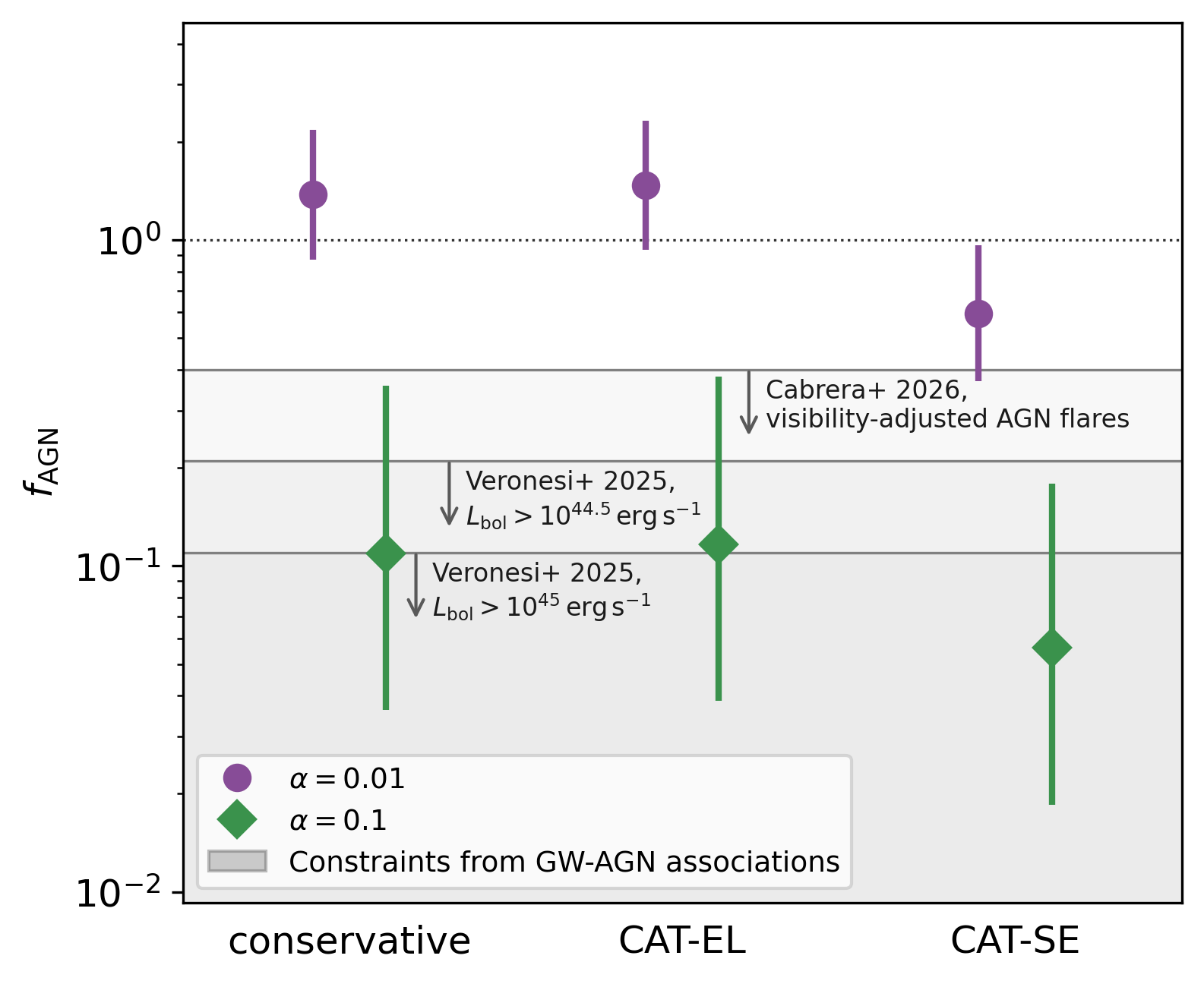}
    \caption{Integrated contribution of the AGN channel to the BBH merger rate density, $f_{\rm AGN}$, computed as the AGN merger rate density for the population with $3\,M_\odot \leq m_2 \leq m_1 \leq 300\,M_\odot$, integrated over the redshift range $0 \leq z \leq 1.5$ and normalized by the corresponding integral of the GWTC-5 DefaultBBH model \citep{GWTC5_pop}. Results are shown for the conservative, CAT-EL, and CAT-SE cosmological AGN population models. Purple circles and green diamonds correspond to disk-viscosity parameters $\alpha=0.01$ and $0.1$, respectively. Symbols indicate the logarithmic mean over the adopted metallicity prescriptions and initial BH orbital dispersions, while the vertical bars span the full range. 
    Gray horizontal lines, arrows, and shaded regions show observational upper limits from GW–AGN associations for reference: the limits of \citet{Veronesi_2025} apply to unobscured AGNs above the indicated bolometric luminosity thresholds, whereas the \citet{Cabrera_2026} limit is obtained by translating ZTF flare constraint using the flare-visibility assumptions of \citet{Graham_2023}.} 
    \label{fig:MRD_integrated_fraction}
\end{figure}

The merger rate densities predicted by our AGN-channel models can be compared with the BBH population inferred from current GW observations. To ensure a consistent comparison, we restrict the ARISE rate to the mass domain of the GWTC-5 DefaultBBH model \citep{GWTC5_pop}. We define
\begin{equation}
\mathcal{R}_{\rm AGN}^{\rm matched}(z)
=
\int_{\mathcal{D}_{\rm LVK}}
\frac{{\rm d}^2\mathcal{R}_{\rm AGN}}
     {{\rm d}m_1\,{\rm d}m_2}
(m_1,m_2,z)\,
{\rm d}m_1\,{\rm d}m_2,
\end{equation}
where 
$\mathcal{D}_{\rm LVK}$ is the domain where $3\,M_\odot \leq m_2 \leq m_1 \leq 300\,M_\odot$. We quantify the AGN-channel contribution through the redshift-integrated rate-density ratio
\begin{equation}
f_{\rm AGN}
=
\frac{
\displaystyle \int_{z_{\rm min}}^{z_{\rm max}}
\mathcal{R}_{\rm AGN}^{\rm matched}(z)\,{\rm d}z
}{
\displaystyle \int_{z_{\rm min}}^{z_{\rm max}}
\mathcal{R}_{\rm LVK}(z)\,{\rm d}z
},
\label{eq:fAGN}
\end{equation}
where \(\mathcal{R}_{\rm LVK}(z)\) is the merger-rate density inferred from the GWTC-5 DefaultBBH population model, integrated over the same mass domain. We integrate over the redshift interval covered by the LVK inference, adopting $z{\rm min}=10^{-6}$ and $z_{\rm max}=1.5$. This quantity provides a rate-based estimate rather than a direct inference of the astrophysical branching fraction.

As shown in \autoref{fig:MRD_integrated_fraction}, the predicted AGN contribution depends most strongly on the assumed disk viscosity. For $\alpha=0.01$, the logarithmic-mean contribution ranges from approximately $f_{\rm AGN}=60\%$ in the CAT-SE model to $f_{\rm AGN}\simeq140$--$150\%$ in the conservative and CAT-EL models, as they overproduce BBH mergers at $z\gtrsim0.4$ with respect to the LVK observations. By contrast, for $\alpha=0.1$, it ranges from approximately $f_{\rm AGN}=6\%$ to $f_{\rm AGN}=12\%$.

The predicted values of $f_{\rm AGN}$ also depend on the cosmological AGN population model, although more weakly. The conservative and CAT-EL prescriptions yield similar integrated fractions, differing by $\lesssim7\%$, whereas the CAT-SE prescription reduces $f_{\rm AGN}$ by a factor of approximately $2$--$2.5$. At fixed $\alpha$ and AGN population model, the initial BH orbital dispersion produces a larger variation than the metallicity prescription. Changing from $\sigma_{\rm BH}=0.0$ to $0.1$ reduces the predicted fraction by a factor of about $2.3$ for $\alpha=0.01$ and by an average factor of about $9$ for $\alpha=0.1$. In comparison, switching between the flat- and evolving-$Z$ models generally changes $f_{\rm AGN}$ by no more than a factor of $1.2$. The individual values for all submodels analyzed in this paper are reported in \autoref{tab:fAGN}.

Additional (although not directly equivalent) constraints on $f_{\rm AGN}$ come from spatial cross-correlation analyses between GW sources and AGN catalogs. \citet{Veronesi_2025} cross-matched the localization volumes of 159 BBH merger events with the Quaia catalog at $z\leq1.5$, obtaining $95\%$ credible upper limits of $f_{\rm AGN}<21\%$ and $<11\%$ for unobscured AGN brighter than $L_{\rm bol}=10^{44.5}$ and $10^{45}{\rm erg/s}$, respectively. 
Complementary searches for transient electromagnetic counterparts provide constraints on the subset of AGN mergers producing detectable electromagnetic emission. 
By correlating 78 BBH mergers with 20 candidate ZTF AGN flares, \citet{Veronesi_2025b} constrained the fraction of detected mergers physically associated with an observed flare to $f_{\rm flare}<15.5\%$ at $90\%$ credibility. Their analysis is primarily sensitive to whether candidate flares are spatially overrepresented within GW localization volumes, while explicitly accounting for the incomplete ZTF coverage of each BBH merger event. Similarly, \citet{Cabrera_2026} analyzed 76 BBHs together with 17 of the same candidate flares using an inhomogeneous Poisson population model that combines the expected number and spatial distribution of both true counterparts and unrelated AGN flares. In this framework, the large number of BBHs without a credible counterpart provides strong leverage against high counterpart fractions, leading to a tighter limit of $f_{\rm flare}<3\%$ at $90\%$ credibility on the fraction of BBH mergers producing observable AGN flares. However, this quantity is not equivalent to the total contribution of the AGN channel, as some BBH mergers occurring in AGN disks may not produce an observable electromagnetic counterpart. Accounting for estimates of the detection probability of an AGN flare associated to a BBH merger \citep{Graham_2023}, \citet{Cabrera_2026} conclude that their limit of $f_{\rm flare}<3\%$ remains compatible with an intrinsic AGN-disk contribution as large as $f_{\rm AGN}\sim40\%$.

We report these estimates in \autoref{fig:MRD_integrated_fraction}, and note that our low-viscosity models ($\alpha=0.01$) systematically exceed the corresponding limits, even in the CAT-SE prescription. Models with $\alpha=0.1$ are instead compatible with these constraints.
A direct comparison would require applying the same selection criteria as \citet{Veronesi_2025} to the simulated AGN hosts and specifying the fraction of BBH mergers that produce a detectable electromagnetic counterpart for flare-based searches \citep{Cabrera_2026}, which we leave for future work.

\subsection{AGN disk lifetime and duty cycle}
\label{sec:discussion_AGN_lifetimes}

The lifetime of an AGN disk, $\tau$, represents an important uncertainty in our model since it sets the time available for BH capture, migration, binary formation, and repeated mergers \citep[see, e.g., Section 4.1 in][]{Vaccaro_2026}. We adopt AGN lifetimes motivated by the quasar lifetime distribution inferred by \citet{Khrykin_quasar_lifetime} from HeII proximity zones. Their analysis is based on 22 quasars at $2.7\lesssim z\lesssim3.9$ and favors a lognormal distribution of quasar lifetimes with mean $ \log{(\tau/{\rm Myr})} = 0.22$ and standard deviation $\sigma_{\tau} = 0.8$. However, the quantity most directly probed by proximity zones is the so-called on-time, $t_{\rm on}$, i.e. the time elapsed since the beginning of the current UV-luminous accretion episode. The underlying total episode duration, $\tau$, is inferred statistically by assuming that the observed quasars are randomly sampled during their accretion episodes, with $t_{\rm on}$ uniformly distributed between zero and $\tau$. 
Consequently, these measurements primarily constrain the duration of individual radiatively bright episodes rather than the cumulative time over which a SMBH may undergo AGN activity.

This distinction is particularly relevant for our interpretation of $\tau$ as the lifetime of the gaseous disk. A luminous quasar episode does not necessarily match the time of existence of a dynamically relevant AGN disk. Accretion may continue in obscured or radiatively inefficient phases that are not captured by proximity-zone measurements, as also discussed by \citet{Khrykin_quasar_lifetime} and \citet{Worseck_2021}. Conversely, AGN activity may be intermittent, with several luminous episodes separated by quiescent intervals. In our calculations, each AGN is instead represented by a single continuous disk episode of duration $\tau$, and any previous or subsequent activity is neglected. For instance, if stellar-mass BHs remain bound to the nuclear region between accretion episodes, their migration and hierarchical merger histories could resume when activity restarts. The cumulative time available for hierarchical growth could then substantially exceed the duration of any individual luminous episode, potentially enhancing the fraction of repeated mergers and significantly modifying the merger delay-time distribution. This effect is not necessarily monotonic, however: its magnitude would depend on whether embedded BHs are retained in the NSC during the inactive phase, as well as on how closely the structure, orientation, and radial extent of a newly formed disk resemble those of the preceding episode.

A further limitation is that we apply the same lifetime prescription independently of SMBH mass, Eddington ratio, and redshift. \citet{Worseck_2021} found no significant correlation between the inferred quasar on-time and either $\MSMBH$ or $\fEdd$. This result should not, however, be interpreted as evidence for a universal lifetime distribution. SMBH masses and Eddington ratios were available only in a relatively restricted region of parameter space, with $\MSMBH\sim10^9$--$10^{10}\Msun$, luminosities broadly consistent with Eddington-limited accretion, and $z\sim3$--$4$. Their results therefore provide little constraint on the lower-mass SMBHs, strongly sub-Eddington systems, nor on possible evolution of the lifetime distribution towards higher redshift.
There are, in fact, physical reasons to expect the duration of an accretion episode to depend on the accretion state. Recently, \citet{Johnson_2026} analytically derived the maximum timescale over which an AGN disk can sustain a given accretion rate before becoming gravitationally unstable. For a disk supported by gas pressure alone, they obtain $\tau_{\rm max}\propto 
\dot{\MSMBH}^{-17/18}$, implying shorter sustainable accretion episodes at higher accretion rates for fixed gas viscosity and SMBH mass. They show that magnetic pressure can stabilize the outer disk and extend the duration of sustained accretion, softening its dependency on the accretion rate to $\tau_{\rm max}\propto 
\dot{\MSMBH}^{-1/2}$. These results reinforce the possibility that both the characteristic AGN lifetime and the underlying disk structure depend on the SMBH mass, accretion state, and magnetic field. In particular, assigning the same lifetime distribution to the Eddington-limited and super-Eddington populations considered here may overestimate the duration of the phases with highest accretion rate, unless additional mechanisms, such as magnetic support, allow these disks to remain stable for longer. A self-consistent distribution $p(\tau \mid \MSMBH,\fEdd,z)$, together with an explicit treatment of recurrent AGN episodes, will therefore be an important extension of the present model.

\subsection{Other caveats}
The results presented in this work rely on several assumptions concerning both the cosmological AGN population and the local physics of BBH formation in AGN disks. These assumptions should be kept in mind when interpreting the predicted merger rate densities and constraints on $f_{\rm AGN}$.

Our treatment of the AGN properties is necessarily simplified. The \CAT{} simulations \citep{Trinca_2022} provide the high-redshift distributions used in this work, but the interpolation to low redshift and the transition across $2\lesssim z\lesssim 4$ are phenomenological. The model does not capture the full diversity of AGN accretion states, including possible correlations of the SMBH mass and Eddington ratio with its duty cycle and its host-galaxy properties. In particular, we assume that the environment supplying compact objects to the AGN disk evolves concurrently with the SMBH, by applying the relation between SMBH and NSC masses from \citet{Graham_Spitler} at all redshifts. In practice, the reservoir of stellar-origin BHs around SMBHs may vary depending on cosmic time, metallicity, and galactic merger history. The AGN lifetimes and duty cycles are major sources of uncertainty, as discussed in Sec.~\ref{sec:discussion_AGN_lifetimes}. Moreover, we keep the SMBH mass and accretion rate, as well as the disk morphology, fixed during the AGN accretion episode. Instead, they are expected to change slowly over the AGN lifetime as BHs are embedded in the disk \citep{Tagawa_2022} and gas is accreted by the SMBH.

Finally, the predicted AGN-channel merger efficiency depends sensitively on uncertain disk physics. Migration torques, the formation of migration traps, the transition between Type I and Type II migration, the gas-hardening prescription, the initial binary orientation, the interaction between embedded binaries and the surrounding gas, disk turbulence, and magnetic fields all affect whether BBHs form and merge efficiently, as discussed in \citet{Vaccaro_2026}.

\section{Summary}
\label{sec:conclusions}

In this work, we connect the formation and hierarchical growth of BBHs in individual AGN disks to the cosmological evolution of the AGN population. To this end, we introduce \arise{} (AGN Rates and Intrinsic Stellar-BH Evolution), a framework that combines the merger yields predicted by semi-analytical AGN-disk simulations, in this case \fastclusterAGN{} \citep{Vaccaro_2026}, with redshift-dependent distributions of AGN abundance, SMBH mass, Eddington ratio, and nuclear metallicity.
We considered three prescriptions for the evolving AGN population: a conservative model based on the observational constraints of \citet{Shen_2020}, and two models based on the \CAT{} semi-analytical simulations \citep{Trinca_2022}, corresponding to Eddington-limited (CAT-EL) and super-Eddington (CAT-SE) SMBH growth. 
This allows us to predict the BBH merger-rate density and the intrinsic merger population across cosmic time, from the local Universe to $z \simeq 10$. 
Our main results can be summarized as follows.

The overall normalization of the AGN-channel merger rate density is controlled primarily by the properties of the AGN disk. In particular, decreasing the viscosity parameter from $\alpha=0.1$ to $0.01$ increases the predicted merger rate density by approximately one order of magnitude over most of the redshift range (see \autoref{fig:MRD(z)}). In the conservative cosmological model, the merger rate density peaks at $z\simeq1.8$ ($1.5$) for $\alpha=0.01$ ($0.1$), reaching
$\sim280\,{\rm Gpc^{-3}\,yr^{-1}}$ for $\alpha=0.01$, whereas the corresponding high-viscosity model reaches only $\sim19\,{\rm Gpc^{-3}\,yr^{-1}}$.

The redshift evolution of the merger rate density is strongly affected by the assumed history of SMBH growth. The conservative model declines rapidly beyond redshift $z\simeq 2$, while the CAT-EL model peaks later, around $z\simeq3.3$ ($3.6$) for $\alpha=0.01$ ($0.1$). In the CAT-SE scenario, the merger rate density instead remains substantial even at much earlier cosmic times: after reaching a maximum around $z\simeq4$, it approaches a broad plateau at $z\gtrsim6$ that persists up to $z\simeq10$. This behavior is not determined by the AGN abundance alone. It reflects the different distributions of SMBH masses and accretion rates in the two considered \CAT{} scenarios. Super-Eddington growth allows a substantial population of SMBHs to enter the region of $(\MSMBH,\fEdd)$ parameter space where BBH production is most efficient already at high redshift, whereas in the CAT-EL model this population develops later in cosmic history. The BBH merger history in AGN disks therefore retains information about the growth history of the SMBH population.

At the redshifts currently probed by LVK ($0 \leq z \leq 1.5$), the predicted contribution of the AGN channel to the BBH merger rate density, $f_\mathrm{AGN}$, is set mainly by the disk viscosity (see \autoref{fig:MRD_integrated_fraction}). For $\alpha = 0.01$, the AGN channel alone gives $f_\mathrm{AGN} \simeq 37$--$96\%$ in the CAT-SE model and $f_\mathrm{AGN} \simeq 87$--$232\%$ in the conservative and CAT-EL models, overproducing the LVK merger rate at $z \gtrsim 0.4$. For $\alpha = 0.1$, the contribution drops to $f_\mathrm{AGN} \simeq 2$--$38\%$. The cosmological AGN model has a weaker effect: the conservative and CAT-EL models agree to within $\sim 7\%$, while CAT-SE lowers $f_\mathrm{AGN}$ by a factor of $\sim 2$. Within each model, the initial BH orbital dispersion $\sigma_\mathrm{BH}$ adds a further uncertainty of a factor of $\sim 2$ for $\alpha = 0.01$ and $\sim 9$ for $\alpha = 0.1$. The high-viscosity models ($\alpha=0.1$) are consistent with current upper limits from GW--AGN spatial cross-correlations ($f_\mathrm{AGN} < 11$--$21\%$; \citealt{Veronesi_2025}) and from searches for AGN flare counterparts ($f_\mathrm{AGN} \lesssim 40\%$; \citealt{Cabrera_2026}). The low-viscosity models exceed both limits in most AGN population model.

The ET and other third-generation detectors will extend the accessible BBH population well beyond the redshift range currently probed by LVK, into the regime where the cosmological models considered here diverge most strongly. For a one-year observing run with duty factor $0.85$, we predict $\sim10^5$ detectable mergers ($\rho>12$) for $\alpha=0.01$, and $\sim 10^4$ for $\alpha=0.1$, with $\sim 10$--$30\%$ of them at $\rho>100$ (see \autoref{tab:gwfish_Nevents}). The total number of detections varies by less than a factor of two between the cosmological models, whereas their redshift distribution differs markedly (see \autoref{fig:gwfish_event_counts}). In the conservative model the detections drop to zero beyond $z\simeq 5$--$6$, and in the CAT-EL model beyond $z\simeq 6$--$7$. In the CAT-SE model, instead, a significant population remains detectable up to $z\simeq 8$. The persistence or disappearance of this high-redshift AGN BBH merger population therefore provides a potential way to distinguish between different histories of SMBH growth.

\section*{Data availability}
The code used in this study is publicly available via GitLab at \href{https://gitlab.com/mariapaola.vaccaro/arise#}{gitlab.com/mariapaola.vaccaro/arise}. Supplementary material is on Zenodo at \href{https://zenodo.org/records/23693189}{zenodo.org/records/23693189}.
The data underlying this article are available upon reasonable request to the corresponding authors, and will be made public upon publication. 

\begin{acknowledgements}
We thank Marta Volonteri, Ralf Klessen, Manuel Arca Sedda, Biswajit Banerjee, Filippo Santoliquido, Marica Branchesi, Andrea Lapi, Ilya Khrykin, Laura Sberna, Jupiter Stevenson, Debora Sijacki, Roberto Maiolino, Yannic Pietschke, and Om Sharan Salafia for useful discussions. We thank Dylan Nelson for granting us access to the IllustrisTNG JupyterLab workspace. 

MPV and MM acknowledge financial support from the European Research Council for the ERC Consolidator grant DEMOBLACK, under contract no. 770017 (PI:~M.~Mapelli).  MM acknowledges support  from the European Research Council for the ERC Advanced grant IMBLACK, under contract no. 101197608 (PI:~M.~Mapelli). MPV and MM acknowledge financial support from the German Excellence Strategy via the Heidelberg Cluster of Excellence (EXC 2181 - 390900948) STRUCTURES. MPV acknowledges support from the Scientific Software Center (SSC) at Heidelberg University through the SSC Fellows program, and is especially grateful to Tuyen Le and 
Dmitrii Kapitan. MPV acknowledges the ACME project, which has received funding from the European Union’s Horizon Europe Research and Innovation programme under Grant Agreement No. 101131928. RS acknowledges support from the MUR projects FIS-2024-01621 DAWN and from the INFN TEONGRAV initiative. 
The authors acknowledge support by the state of Baden-W\"urttemberg through bwHPC and the German Research Foundation (DFG) through grants INST 35/1597-1 FUGG and INST 35/1503-1 FUGG.

This research made use of \href{https://gitlab.com/mariapaola.vaccaro/arise}{\arise{}}, \href{https://gitlab.com/mariapaola.vaccaro/fastcluster_agn}{\fastclusterAGN{}} \citep{Vaccaro_2023, Vaccaro_2026}, \href{https://gitlab.com/micmap/fastcluster_open}{\fastcluster{}} \citep{fastcluster2021}, \href{https://gitlab.com/sevncodes/sevn}{\sevn{}} \citep{sevn_2023}, \href{https://gitlab.com/aatrani/pagn}{\pagn{}} \citep{Gangardt_2024}, \href{https://github.com/janosch314/GWFish}{\gwfish{}} \citep{Dupletsa_2023},
{\fontfamily{lmtt}\selectfont NumPy} \citep{Harris20}, {\fontfamily{lmtt}\selectfont SciPy} \citep{SciPy2020}, {\fontfamily{lmtt}\selectfont AstroPy} \citep{astropy}, {\fontfamily{lmtt}\selectfont Pandas} \citep{Pandas2024}. Plots were produced using {\fontfamily{lmtt}\selectfont Matplotlib} \citep{Hunter2007}.
\end{acknowledgements}

\bibliographystyle{aa} 
\bibliography{bibliography.bib}

\clearpage
\appendix

\refstepcounter{section}
\label{app:review_AGN_obs}

\setcounter{figure}{0}
\renewcommand{\thefigure}{B.\arabic{figure}}

\begin{strip}
    \centering
    \includegraphics[width=\textwidth]
        {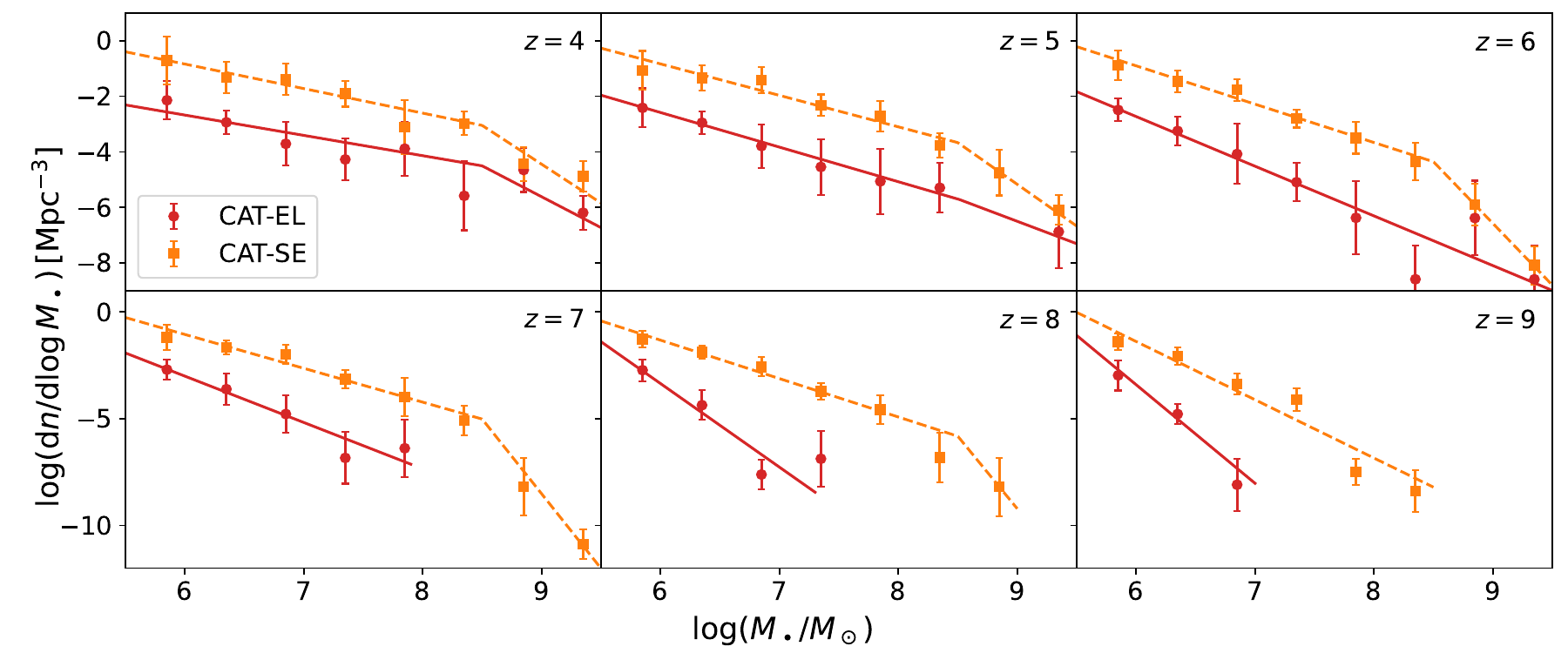}

    \captionof{figure}{
        Black-hole mass functions predicted by the CAT Eddington-limited
        (CAT-EL; red circles) and super-Eddington (CAT-SE; orange squares)
        models at redshifts $z=4$--$9$, as in \citet{Trinca_2022}.
        Solid red and dashed orange curves show the corresponding best-fitting
        broken power laws for each model and redshift.
    }
    \label{fig:BHMF_CAT_data}
\end{strip}

\section*{Appendix \thesection: An evolving picture of AGNs}

Observational constraints on the AGN number density come from surveys across multiple wavelength regimes, including the optical, UV, IR, and X-ray, each affected by different selection effects and levels of obscuration \citep[e.g.][]{Ueda_2014, Aird_2015, Buchner_2015, Miyaji_2015}. A comprehensive synthesis at redshift $0\leq z \leq 7$ was presented by \citet{Shen_2020}, who performed a simultaneous fit to the available measurements at the time, combining multi-band luminosity functions with completeness corrections. Most cosmological simulations tend to overpredict the AGN number density relative to these observational constraints, as shown by \citet{Habouzit_2022, Habouzit_2022b}. 

More recently, however, JWST observations have provided support for a substantially higher AGN number density at high redshift than previously inferred. They revealed a population of compact red sources at high redshift, commonly referred to as little red dots (LRDs). Some LRDs exhibit broad emission lines consistent with an AGN contribution \citep[e.g.][]{Greene_2024, Greene_2026,degraaff_2026}, potentially implying a substantially higher AGN number density at $4\lesssim z \lesssim9$ than previously inferred.
Nevertheless, the physical nature of LRDs remains uncertain. Possible interpretations include massive galaxies \citep{Labbe_2023} and proto-globular clusters \citep{Chisholm_2026}, which are not necessarily mutually exclusive with the AGN interpretation \citep{Liempi_2026}, but also supermassive stars \citep{Zwick_2025, Nandal_2026, Tanikawa_2026} and quasi-stars \citep{Umeda_2025, Begelman_2026, Santarelli_2026, Gentile_2026}.

Evidence for a high abundance of active SMBHs in the early Universe does not rely exclusively on LRDs. For instance, \citet{Maiolino_2024} and \citet{Taylor_2025} identify AGN candidates through resolved broad emission lines; these samples are not selected as LRDs, although they may contain a limited contribution from objects that would also satisfy LRD classifications. \citet{Geris_2026} instead stack the spectra of a more general population of JADES galaxies, searching for evidence of broad-line components that cannot be identified in individual spectra, finding that the AGN number density at higher redshift, $z\gtrsim4$, appear to be broadly consistent with estimates obtained from cosmological simulations. Independently, \citet{Scholtz_2025} select Type-II AGNs through high-ionization emission lines. Complementary evidence comes from deep JWST mid-infrared imaging and spectral energy distribution (SED) modeling, which \citet{Bulichi_2026} use to construct the total AGN bolometric luminosity function over $0.5<z<6$. They infer an obscured fraction that increases strongly with redshift, exceeding approximately $90\%$ at $z\gtrsim2.5$ and reaching $\sim98\%$ in their highest-redshift bin. At redshift $z\gtrsim2.5$, their luminosity function lies well above UV- and X-ray-based estimates, including \citet{Shen_2020}, suggesting that previous surveys may miss a substantial population of heavily obscured AGNs.
Although each of these samples is subject to different selection effects and possible contaminants, their number-density estimates collectively support the possibility of a numerous population of active SMBHs at high redshift.

\section{The SMBH mass distribution}
\label{app:M_SMBH_z}

\begin{figure*}[!p]
    \centering
    \setcounter{figure}{1}
    \renewcommand{\thefigure}{B.\arabic{figure}}
    
    \begin{minipage}[t]{0.48\textwidth}
        \centering
        \includegraphics[width=0.98\linewidth]
            {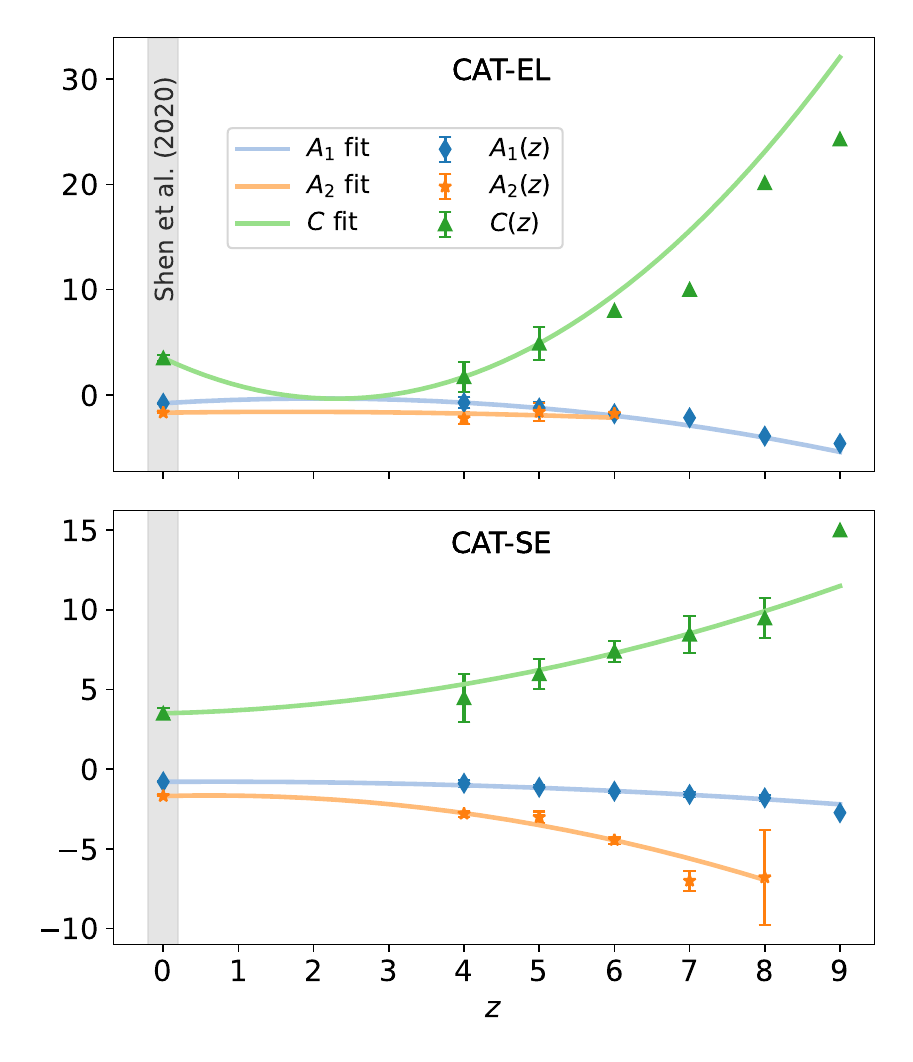}
        \captionof{figure}{Redshift evolution of the parameters of the broken-power-law fits to the black-hole mass function for the CAT-EL (top) and CAT-SE (bottom) models. Diamonds, stars, and triangles show the best-fitting values of the low-mass slope $A_1$, high-mass slope $A_2$, and intercept $C$, respectively; error bars indicate the corresponding $1\sigma$ uncertainties. Continuous lines show the quadratic redshift-evolution fits, anchored at $z=0$ to the fit obtained from the \citet{Shen_2020} mass function.}
        \label{fig:BHMF_parameters_vs_z}
    \end{minipage}
    \hfill
    \begin{minipage}[t]{0.48\textwidth}
        \centering
        \includegraphics[width=\linewidth]
            {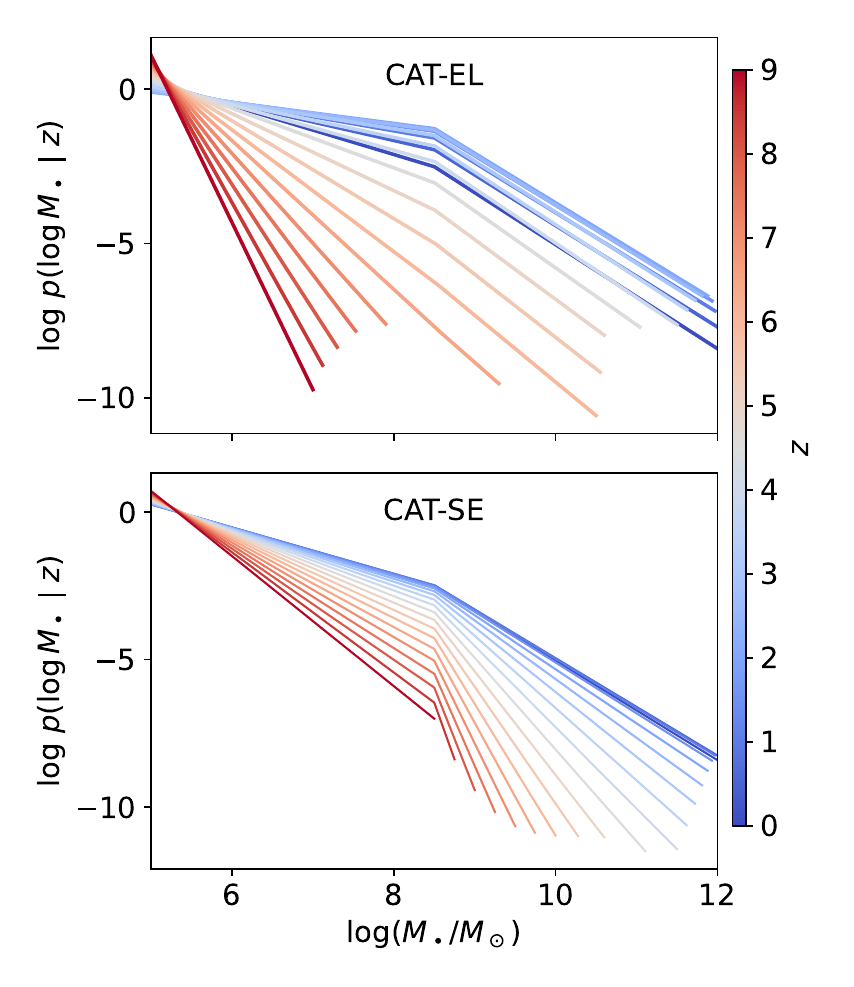}
        \captionof{figure}{Normalized SMBH-mass probability distributions for the CAT-EL (top) and CAT-SE (bottom) models. Curves show the distributions at redshifts $z=0$--$9$, color-coded as indicated by the color bar, obtained by normalizing the reconstructed SMBH mass functions according to Eq.~\ref{eq:pM_norm}. Each distribution is truncated at the corresponding redshift-dependent maximum mass, $\mu_{\max}(z)$. }
        \label{fig:BHMF_allz}
    \end{minipage}

    \vspace{1cm}

    \setcounter{figure}{0}
    \renewcommand{\thefigure}{C.\arabic{figure}}
    
    \begin{minipage}{\textwidth}
        \centering
    \includegraphics[width=\linewidth]{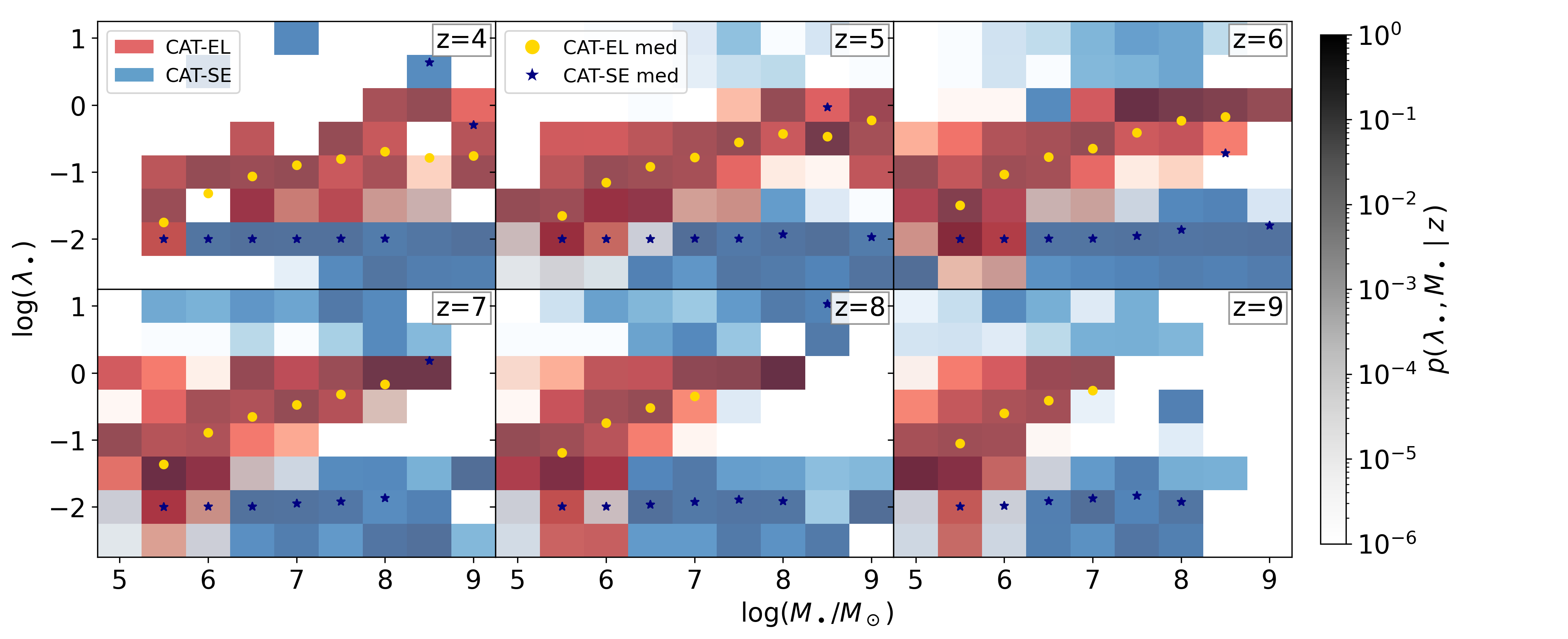}
    \captionof{figure}{Joint distribution of SMBH mass, $\MSMBH$, and Eddington ratio, $\fEdd$, predicted by the CAT-EL (red) and CAT-SE (blue) models at $z=4$--$9$. The color intensity indicates the probability density $p(\fEdd,\MSMBH|z)$ in each bin; overlapping regions appear purple. Yellow circles and blue stars mark the median $\fEdd$ at fixed $\MSMBH$ for the CAT-EL and CAT-SE models, respectively.}
    \label{fig:pfedd_CAT_comparison}
    \end{minipage}

\end{figure*}

To assign SMBH masses as a function of redshift, we construct a normalized probability distribution,
$p(\MSMBH|z)$, from redshift-dependent SMBH mass functions. We use the \CAT{} SMBH mass functions of
\citet{Trinca_2022} at high redshift, $z\geq 4$, considering both the Eddington-limited and super-Eddington growth
scenarios, hereafter CAT-EL and CAT-SE. The tabulated quantities are given as comoving number densities per
logarithmic SMBH-mass interval, $\Phi(\MSMBH,z) = dn / d\mu$,
on a grid in $\mu \equiv \log (\MSMBH/\Msun)$.

We fit each tabulated mass function in each redshift snapshot with a continuous broken power law in log-space, namely
\begin{equation}
\log\Phi =
\begin{cases}
A_1\,\mu + C, & \mu_{\rm min}\leq \mu \leq \mu_{\rm b} \\
\left[A_1-A_2\right]\mu_{\rm b}
+ A_2\,\mu + C, &  \mu_{\rm b} < \mu \leq \mu_{\rm max}
\end{cases}
\label{eq:bhmf_broken_powerlaw}
\end{equation}
as shown in \autoref{fig:BHMF_CAT_data}. The parameters $A_1$, $A_2$, and $C$ describe the low-mass slope, high-mass slope, and normalization of the mass function, respectively. The best fit functions for $A_1(z_i), A_2(z_i), C(z_i)$ at the tabulated redshifts $z_i=4,5,6,7,8,9$ are shown in \autoref{fig:BHMF_parameters_vs_z}. We adopt a minimum mass $\mu_{\rm min}=5$ and a break mass $\mu_{\rm b}=8.5$.
The maximum mass, $\mu_{\rm max}$, is redshift-dependent and is chosen so as not to sample SMBH masses beyond the region supported by data. For the CAT-EL (CAT-SE) scenario, we use $\mu_{\rm max} = 11.5,10.6,10.5,7.9,7.3,7.0$ ($11.5,10.6,10.0,9.5,9.0,8.5$)
at $z = 4,5,6,7,8,9$ respectively.

We anchor the low-redshift distribution
using the local SMBH mass function inferred by \citet{Shen_2020}. We use their deconvolved local SMBH mass function, which is intended to represent the intrinsic SMBH population rather
than the observed distribution. 
We perform the same fit as in Eq.~\ref{eq:bhmf_broken_powerlaw} for the \citet{Shen_2020} mass function at $z\simeq 0$, obtaining $A_1(z=0)$, $A_2(z=0)$, and $C(z=0)$.

We then model the redshift evolution of each broken-power-law parameter with a quadratic function,
\begin{equation}
X(z) = X_0 + \beta_X z + \alpha_X z^2,
\label{eq:anchored_quadratic_bhmf}
\end{equation}
where $X \in {A_1,A_2,C}$, and $X_0\equiv X(z=0)$ is fixed by the \citet{Shen_2020} local anchor. The free parameters $\beta_X$ and $\alpha_X$ are obtained via a fitting procedure, separately for CAT-EL and CAT-SE, as shown in \autoref{fig:BHMF_parameters_vs_z}.
Thus, the resulting SMBH mass functions are smooth parametric reconstructions anchored to the local \citet{Shen_2020} mass function at $z=0$ and calibrated to the \CAT{} mass functions at high redshift.

For each available redshift slice, we convert the SMBH mass function into a probability distribution by
normalizing over the mass range used in the AGN population model,
\begin{equation}
p(\mu|z) =
\frac{\Phi(\mu|z)}
{\int_{\mu_{\rm min}}^{\mu_{\rm max}(z)}
\Phi(\mu|z)\,d\mu},
\label{eq:pM_norm}
\end{equation}
obtaining the distributions shown in \autoref{fig:BHMF_allz}.

\vspace{0.5cm}

\section{The Eddington-ratio distributions}
\label{app:fEdd_M_z}

We require a distribution of Eddington ratios, $p(\fEdd|\MSMBH,z)$, to assign the accretion state of each AGN host. The fiducial low-redshift prescription is
based on the Eddington-ratio distribution adopted by \citet{Shen_2020}, following the form introduced by
\citet{Tucci_Volonteri_2017}. This distribution is written as the sum of two components,
\begin{align}
p_{\rm Shen}( \fEdd | z) &= 
(1-F)\,A\,\fEdd^{1+a}\exp\left(-\fEdd / \lambda_1\right)
+ \\ &+
\frac{F}{\sqrt{2\pi\sigma^2}}
\,\exp\left[
-\frac{
\left(\log\fEdd-\log\lambda_2\right)^2
}{2\sigma^2}
\right], \nonumber
\label{eq:shen_erdf}
\end{align}
where the first term describes the obscured/type-2 AGN population and the second term describes the
unobscured/type-1 AGN population. We adopt the parameters of \citet{Shen_2020}, with
$a=-0.6$, $\lambda_1=1.5$, and a type-1 fraction $F=0.38$. The lognormal component evolves with redshift according
to
\begin{align}
\log\lambda_2(z)
&=
\max\left[
-1.9+0.45 z,\,
\log(0.03)
\right],\\
\sigma &= 
\left(
1.03+0.15 z
\right)/ \ln 10 .
\label{eq:shen_lambda2}
\end{align}

\setcounter{figure}{1}
\renewcommand{\thefigure}{C.\arabic{figure}}

\begin{figure}
    \centering
    \includegraphics[width=\linewidth]{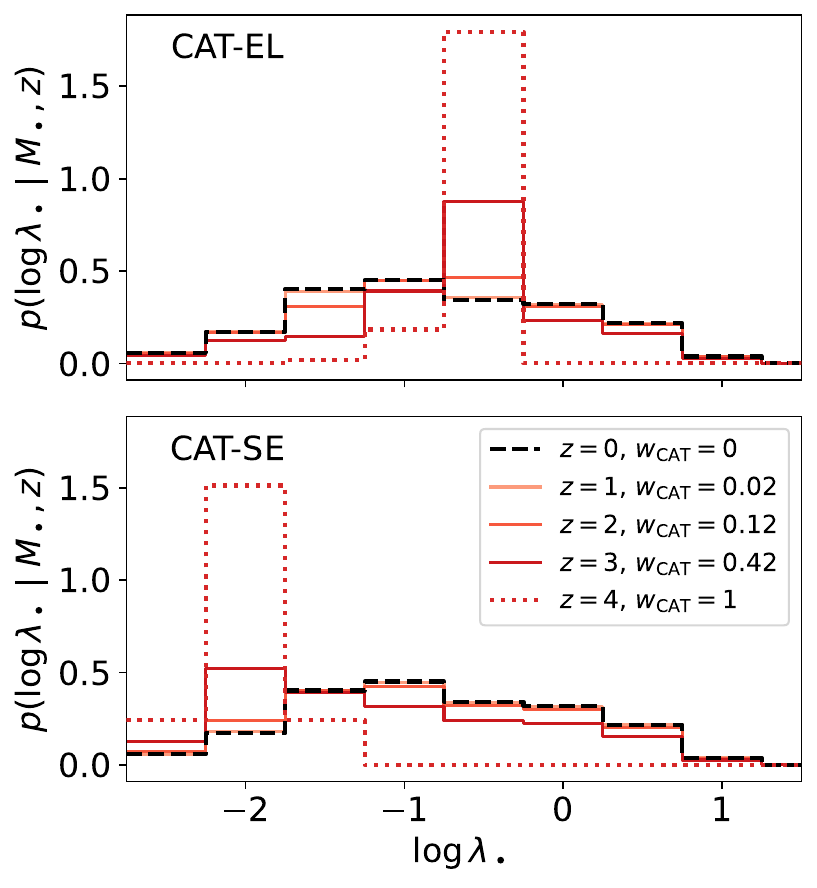}
    \caption{Illustration of the adopted low-redshift bridging procedure for the Eddington-ratio distribution in the CAT-EL (top panel) and CAT-SE (bottom panel) models, at fixed SMBH mass $\log_{10}(M_\bullet/M_\odot)=7.5$. The dashed black curve shows the mass-independent \citet{Shen_2020} distribution adopted at $z=0$, while the red dotted curves show the corresponding \CAT{} distributions at $z=4$. Thinner continuous curves show the interpolated distributions at $z=1$, $2$, and $3$, obtained using the redshift-dependent weight $w_{\rm CAT}(z)=(z/4)^3$.}
    \label{fig:pfedd_bridge_EL_SE}
\end{figure}

At high redshift, $z\geq 4$, the \CAT{} simulations provide probability distributions of the Eddington ratio as a function of SMBH mass and redshift, $p_{\rm CAT}(\lambda|\MSMBH,z)$ for both the CAT-EL and CAT-SE models, shown in \autoref{fig:pfedd_CAT_comparison}. For $z\geq 4$, we use the \CAT{} tables directly at integer $z$ and interpolate in
log-probability space between
the two nearest \CAT{} snapshots at non-integer redshifts.

For intermediate redshift, $0\lesssim z < 4$, we construct a nonlinear bridge between the \citet{Shen_2020} distribution and
the \CAT{} $z=4$ distributions, namely
\begin{align}
p(\fEdd|\MSMBH,z)
&=
\left[1-w_{\rm CAT}(z)\right]\,p_{\rm Shen}(\fEdd|z)
+ \\ &+
w_{\rm CAT}(z)\,
p_{\rm CAT}(\fEdd|\MSMBH,z=4), \nonumber
\label{eq:pfedd_bridge}
\end{align}
where
$w_{\rm CAT}(z)=\left(z / 4\right)^\gamma$ and $\gamma=3$. 
The resulting distributions, of which an example is shown in \autoref{fig:pfedd_bridge_EL_SE}, are
re-normalized over $\fEdd$ at every $(\MSMBH,z)$, such that
$\int d\fEdd\, p(\fEdd|\MSMBH,z) = 1$ over the considered range $\fEdd \in [10^{-3}, 10]$.

\section{Number of black holes}
\label{app:N_BH}

We compute the approximate number of stellar-mass BHs,
$N_{\rm BH}$, expected to interact with the disk of an AGN
powered by an SMBH of mass $\MSMBH$ as in \citet{Vaccaro_2023}.

We assume that the SMBH coexists with a NSC. Following \citet{Graham_Spitler},
we adopt the approximate relation
\begin{equation}
    \log{\tonde{\frac{\MSMBH}{\MSMBH+M_\mathrm{NSC}}}} =
    \frac{2}{3}\, \log{\tonde{\frac{\MSMBH}{5\times10^7\, \Msun}}}
    \label{eq: SMBH-NSC-Graham-Spitler}
\end{equation}
for $\MSMBH \leq 4.3\times10^7\,\Msun$, and $\MSMBH/(\MSMBH+M_\mathrm{NSC}) = 0.9$
otherwise. 
We draw the NSC effective radius from the mass--radius relation of
\citet{Neumayer_2020}, including its scatter, and describe the stellar
distribution with a \citet{BW} cusp. The number of BHs in the disk is then
\begin{equation}
    N_\mathrm{BH}=f_\mathrm{BH}\, \frac{M_{\rm NSC}(R_\mathrm{max})}{\average{m_\ast}}\,,
    \label{eq:N_BH_max}
\end{equation}
where $M_{\rm NSC}(R_\mathrm{max})$ is the NSC mass enclosed within the outer
disk radius $R_\mathrm{max}=0.1\, {\rm pc}\,{}(\MSMBH/10^6\Msun)^{1/2}$ \citep{Goodman_2003, Yang_2019_bis}, $f_\mathrm{BH}\sim\mathcal{N}(0.04,\,0.01)$ is the
BH fraction in the NSC \citep{Bartos_2017}, and $\average{m_\ast}$ is the mean stellar mass of a
\citet{Kroupa_2001} initial mass function.

\section{Population variants}
\label{app:population_variants}

\setcounter{figure}{0}
\renewcommand{\thefigure}{E.\arabic{figure}}

\begin{figure}
    \centering
    \includegraphics[width=\linewidth]{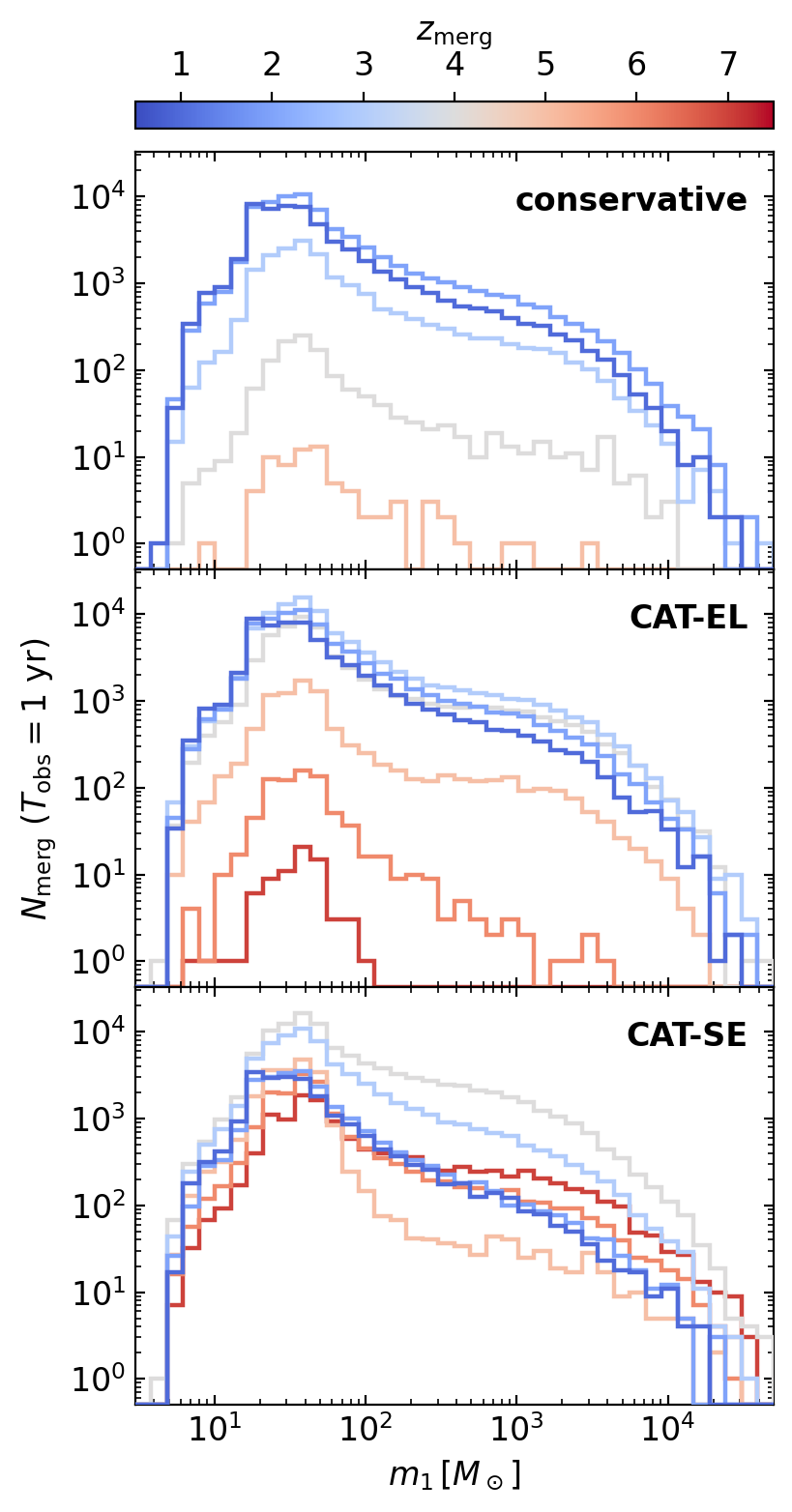}
    \caption{Merger-count distributions of primary mass, $m_1$, for three cosmological model variants: conservative (top row), CAT-EL (middle row), and CAT-SE (bottom row). We assume $\alpha=0.01$, $\sigma_{\rm BH}=0.0$, and the evolving-$Z$ distribution. Each curve shows the number of intrinsic mergers expected over an observing time of $T_{\rm obs}=1\,{\rm yr}$, obtained by integrating the observer-frame merger rate over redshift bins of width $\delta z = 0.5$. Colors encode the merger redshift, $z_{\rm merg}$, increasing from low redshift (blue) to high redshift (red). The corresponding distributions for the flat-$Z$
    model are shown in the first column of \autoref{fig:redshift_colored_distributions}.}
    \label{fig:fig_pm1_evolving_Z}
\end{figure}

In Sec.~\ref{sec:results_intrinsic_population} we discussed the redshift evolution of the
intrinsic BBH population for a representative configuration with
$\alpha = 0.01$, $\sigma_{\rm BH} = 0.0$, and the flat-$Z$ prescription. Here we show the limited impact of the assumed metallicity evolution of galactic nuclei (Sec.~\ref{sec:metallicity_models}). \autoref{fig:fig_pm1_evolving_Z} shows the merger-count distributions of the
primary mass, $m_1$, for the same AGN-disk configuration as \autoref{fig:redshift_colored_distributions}, but adopting the
evolving-$Z$ model. Because the two metallicity prescriptions share the same
AGN abundance, $n_{\rm AGN}(z)$, SMBH mass distribution, $p(M_\bullet\,|\,z)$,
and Eddington-ratio distribution, $p(\lambda_\bullet\,|\,M_\bullet, z)$, any
difference between the two figures arises solely from the metallicity of the
stellar populations in which first-generation (1g) BHs form.
 
The number of mergers in each redshift bin, and its evolution across the
conservative, CAT-EL, and CAT-SE prescriptions, is almost unaffected by the
metallicity model.  
The metallicity prescription instead shapes the low-mass end of the
distribution, which is dominated by first-generation primaries. In the flat-$Z$ model, the
primary-mass distribution peaks at $m_1 \approx 15$--$20\,{\rm M}_\odot$ in
all redshift bins, as expected for stellar populations of approximately solar
metallicity. In the evolving-$Z$ model, the peak moves to
$m_1 \approx 30$--$40\,{\rm M}_\odot$ and broadens because metal-poor progenitors lose less mass through stellar winds and
can therefore collapse into more massive BHs \citep{sevn_2023}. 
 
Above $m_1 \sim 10^2\,{\rm M}_\odot$, where the population is dominated by
higher-generation primaries, the two metallicity prescriptions yield
similar distributions: the hierarchical tail has a comparable slope and
extends to $m_1 \gtrsim10^4\,\Msun$ in the most populated
redshift bins. 
Once BHs have grown through several generations, their masses are set mainly
by the number of mergers they undergo, which depends on the disk properties
and on the SMBH mass and Eddington ratio of the host AGN
\citep{Vaccaro_2026}, rather than by the masses of their $1g$ progenitors. The
assumed metallicity evolution of galactic nuclei therefore mostly affects the
mass range $m_1 \lesssim 10^2\,{\rm M}_\odot$. We note that, in the
evolving-$Z$ model, the $1g$ peak lies close to the feature at
$m_1 \sim 35\,\Msun$ identified in the LVK population
\citep{GWTC5_pop}, whereas the flat-$Z$ model places it at lower masses.
A meaningful comparison, however, would require accounting for the
contribution of other formation channels and for detector selection effects.
 
Variations of \autoref{fig:fig_pm1_evolving_Z} for the secondary mass, the spin parameters
$\chi_{\rm eff}$ and $\chi_{\rm p}$, and the other combinations of $\alpha$
and $\sigma_{\rm BH}$ are available in the supplementary material.

\setcounter{table}{1}
\renewcommand{\thetable}{\arabic{table}}
\begin{table}[]
    \centering
\begin{tabular}{llllr}
\hline\hline
AGN pop. model & $\alpha$ & $\sigma_{\rm BH}$ & $Z$ model & $f_{\rm AGN}$ ($\%$) \\
\hline
conservative & 0.01 & 0.0 & flat-$Z$ & 193 \\
conservative & 0.01 & 0.0 & evolving-$Z$ & 218 \\
conservative & 0.01 & 0.1 & flat-$Z$ & 87 \\
conservative & 0.01 & 0.1 & evolving-$Z$ & 99 \\
conservative & 0.1 & 0.0 & flat-$Z$ & 29 \\
conservative & 0.1 & 0.0 & evolving-$Z$ & 36 \\
conservative & 0.1 & 0.1 & flat-$Z$ & 3.8 \\
conservative & 0.1 & 0.1 & evolving-$Z$ & 3.6 \\
CAT-EL & 0.01 & 0.0 & flat-$Z$ & 206 \\
CAT-EL & 0.01 & 0.0 & evolving-$Z$ & 232 \\
CAT-EL & 0.01 & 0.1 & flat-$Z$ & 93 \\
CAT-EL & 0.01 & 0.1 & evolving-$Z$ & 105 \\
CAT-EL & 0.1 & 0.0 & flat-$Z$ & 31 \\
CAT-EL & 0.1 & 0.0 & evolving-$Z$ & 38 \\
CAT-EL & 0.1 & 0.1 & flat-$Z$ & 4.0 \\
CAT-EL & 0.1 & 0.1 & evolving-$Z$ & 3.9 \\
CAT-SE & 0.01 & 0.0 & flat-$Z$ & 88 \\
CAT-SE & 0.01 & 0.0 & evolving-$Z$ & 96 \\
CAT-SE & 0.01 & 0.1 & flat-$Z$ & 37 \\
CAT-SE & 0.01 & 0.1 & evolving-$Z$ & 40 \\
CAT-SE & 0.1 & 0.0 & flat-$Z$ & 15 \\
CAT-SE & 0.1 & 0.0 & evolving-$Z$ & 18 \\
CAT-SE & 0.1 & 0.1 & flat-$Z$ & 2.0 \\
CAT-SE & 0.1 & 0.1 & evolving-$Z$ & 1.9 \\
\hline
\end{tabular}
    \caption{Integrated AGN contribution, $f_{\rm AGN}$, for each combination of AGN population model, disk-viscosity parameter $\alpha$, initial BH orbital dispersion $\sigma_{\rm BH}$, and metallicity prescription. The values are obtained as in Eq.~\ref{eq:fAGN} and are expressed as percentages.}
    \label{tab:fAGN}
\end{table}

\end{document}